# GitHub Marketplace for Practitioners and Researchers to Date

## A Systematic Analysis of the Knowledge Mobilization Gap in Open Source Software Automation

**Sk Golam Saroar · Waseefa Ahmed · Maleknaz Nayebi**

**Abstract** Marketplaces for distributing software products and services have been getting increasing popularity. GitHub, which is most known for its version control functionality through Git, launched its own marketplace in 2017. GitHub Marketplace hosts third party apps and actions to automate workflows in software teams. Currently, this marketplace hosts 440 Apps and 7,878 Actions across 32 different categories. Overall, 419 Third party developers released their apps on this platform which 111 distinct customers adopted. The popularity and accessibility of GitHub projects have made this platform and the projects hosted on it one of the most frequent subjects for experimentation in the software engineering research. A simple Google Scholar search shows that 24,100 Research papers have discussed GitHub within the Software Engineering field since 2017, but none have looked into the marketplace. The GitHub Marketplace provides a unique source of information on the tools used by the practitioners in the Open Source Software (OSS) ecosystem for automating their project's workflow. In this study, we (i) mine and provide a descriptive overview of the GitHub Marketplace, (ii) perform a systematic mapping of research studies in automation for open source software, and (iii) compare the state of the art with the state of the practice on the automation tools. We conclude the paper by discussing the potential of GitHub Marketplace for knowledge mobilization and collaboration within the field. This is the first study on the GitHub Marketplace in the field.

**Keywords** Software Engineering, GitHub Marketplace, Open Source, Software Automation, Systematic Mapping Study, Knowledge Mobilization

Sk Golam Saroar
York University
E-mail: saroar@yorku.ca

Waseefa Ahmed
York University
E-mail: waseefa@yorku.ca

Maleknaz Nayebi
York University
E-mail: mnayebi@yorku.ca

## 1 Introduction

GitHub Marketplace provides software teams with tools and technologies to improve their development workflows. Third-party developers provide these automation tools on the marketplace to reduce the number of repetitive tasks within a workflow and increase the productivity of a software team. The software engineering research community has also been concerned with the automation of development workflows and providing decision support to assist software teams in improving teams' productivity and products value. Yet, the GitHub Marketplace has never been the subject of an investigation to evaluate the alignment or difference between research efforts (the published studies) and the practice (GitHub automation tools).

Software Marketplaces have been studied since 2012 from the engineering perspective. Harman et al. [16] provided an insight into the importance of marketplace ecosystem for software engineering practices by investigating the mobile app stores. Martin et al. [29] synthesize the research and development directions in mobile app stores. Mobile app stores serve not only as huge collections of apps but also enable developers to produce and distribute content while establishing a communication channel between users and developers via ratings and reviews [29]. The results of their synthesis in 2017 showed that the research in mobile app stores covers the topics of extracting requirements and analyzing user feedback, release and distribution of apps, security and vulnerabilities, and API analysis, as well as the discussion about the ecosystem and the marketplace. Similar to mobile app stores, apps and actions on the GitHub Marketplace also have technical and non-technical attributes.

GitHub Marketplace is a software repository that allows developers to provide free or paid tools to GitHub users for automating their workflows[1], in particular, when hosted on a GitHub repository. Apps and Actions are the two types of tools in this marketplace. While both provide ways to build automation and workflow tools, they each have strengths[2] that make them useful in different ways. **GitHub actions** are designed for automation of customized individual tasks within a workflow [3] including the automation for CI pipelines, building, testing or deploying software applications. They are written in YAML files and can be built as a Docker container or a Javascript snippet. Github actions are free for users. **GitHub apps** are applications that GitHub users can install on user or organization accounts. Apps can access certain repositories when explicitly specified and allowed by the user[4]. Apps can be free or paid, and often each app has multiple paywalls depending on its functionality. GitHub apps are indeed actions that are hosted and run by GitHub. It is recommended that users and organizations use apps if the desired action needs extended permission, or if the user wishes to manage and run the action by themselves, or when users need further architecture and memory support for analyzing data.

Considering the widespread popularity of GitHub in regards to hosting software repositories and managing product releases and the number of tools for workflow automation, we believe that the provided set of tools is highly relevant to real-world software engineering challenges. Hence, the tools are a unique source of data and provide the opportunity to observe practices and understand pain points and trends in software automation. We compared research studies published in the software engineering community and the automation tools devised by the researchers, and focused on studying the gap between research and practice in open source software tools.

[1] https://docs.GitHub.com/en/developers/GitHub-marketplace/GitHub-marketplace-overview/about-GitHub-marketplace

[2] https://docs.github.com/en/actions/creating-actions/about-custom-actions#strengths-of-github-actions-and-github-apps

[3] https://docs.GitHub.com/en/actions/creating-actions/about-actions

[4] https://docs.GitHub.com/en/developers/apps/getting-started-with-apps/about-apps

To the best of our knowledge, this study is the first on the analysis and mining of GitHub marketplace. Our study consists of three main parts. First, we performed an exploratory analysis to understand and evaluate the GitHub Marketplace by mining and synthesizing the technical and non-technical attributes of apps and actions in this ecosystem. Second, we systematically gathered literature on automation tools and techniques. Finally, we mapped state of the art to the state of practice, compared the existing research direction with the automation tools in the GitHub Marketplace, and performed a gap analysis. In particular, we are answering three research questions (RQs):

**RQ1** (Descriptive) **-** What are the characteristics of the GitHub Marketplace and what tools are offered for automating development workflows?
**Why and How:** To provide a better understanding of the attributes and data structure within this repository, we mined the characteristics of apps and actions in the GitHub Marketplace and explained the status quo using all the available fields in the repository and performed a comparative study wherever appropriate.

**RQ2** (Descriptive) **-** How have the automation tools for open source software evolved in the research community over the years?
**Why and How:** We systematically gathered literature on automation tools and techniques within the open source ecosystem and compared the papers over the years. Further, we mapped the topic of these academic studies into the marketplace tool categories to enable comparison between research and current practice.

**RQ3** (Diagnostic) **-** How do the automation tools in the marketplace compare with software engineering literature and what are the gaps?
**Why and How:** We compared the state of the art with the state of the practice. To this end, we first gathered apps and actions information and extracted a list of categories and frequency of coverage from the GitHub Marketplace. We then performed a systematic literature analysis and mapped the papers retrieved from our systematic search into the marketplace categories and discussed the gap and trends.

The rest of this paper is organized as follows. Section 2 presents the related work. Section 3 describes the research methodology, including data extraction, synthesis, and analysis. Section 4 presents the observations from mining GitHub Marketplace. Section 5 shows results of the systematic mapping study. Section 6 provides the gap analysis. Section 7 discusses the threats to validity. Section 8 provides discussion and future implications. Section 9 concludes this paper.

## 2 Related Work

So far, the analysis of marketplaces in software engineering has been limited to mobile apps store analysis. However, marketplaces, and in particular, the GitHub Marketplace has been studied in fields other than the software engineering community from the economical and management points of view.

### 2.1 GitHub Marketplace

Kallis et al. [19, 20] proposed a GitHub app called Issue Tagger which uses machine learning to analyze the title and textual description of issues and automatically assign labels such as bug report, feature request, and question. Experimenting on about 30,000 GitHub issues, they showed that the Ticket Tagger could identify the correct labels with reasonably high effectiveness. They also presented the tool's architecture and provided all the necessary information for installing and using the app on custom GitHub repositories. To improve the quality of software, Souza et al. [44] proposed to analyze code inspection tools available on the Github Marketplace. They selected four tools for the PHP programming language to inspect the GLPI system and found more than ten thousand failures. Subsequently, they analyzed the individual feedback for that tool. Kinsman et al. [23] examined how actions were used by developers and how various activity metrics changed as a result of their adoption. They showed that the use of GitHub actions raised the monthly total of rejected pull requests while lowering the monthly total of commits on merged pull requests. They also found that developers had a favourable opinion of GitHub actions although the technology had only been used by a small percentage of projects thus far. Their findings are especially important for practitioners to understand and prevent undesirable effects on their projects.

There are also multi-sided marketplaces such as Uber, where the platforms have customers on both the demand side and the supply side. Mehrotra et al. [30] presented some research problems in developing a recommendation framework for such marketplaces, for example, introducing a multi-objective ranking that optimizes the different objectives for multi-stakeholders.

## 2.2 Software Bots

Many projects use bots to automate a wide range of tasks such as refactoring source code [51], fixing bugs [33], updating project dependencies [31], supporting communication and decision-making [45], etc. Wessel et. al investigated the usage and impact of such bots in [48]. They classified bots and collected metrics from before and after bot adoption in 93 projects on GitHub. Although the surveyed developers believed that bots were useful for maintenance tasks, the authors found no consistent and statistically significant difference between before and after bot adoption in terms of number of comments, commits, changed files, and time to close pull requests. They also asked developers about their desired improvements for bots. Developers wanted smarter bots that would decrease code review effort, decrease time to close pull-requests, automate continuous integration tasks, among others. Many of the apps and actions in the GitHub Marketplace today automate these tasks. 'Continuous integration' (#1), and 'Code review' (#6) are among the most prevalent topics for actions in the Marketplace (see Table 8).

To assist students in checking code style and functioning, Hu et al. [17] built up a static code analyzer and a continuous integration service on GitHub. The authors implemented three bots and showed that more than 70% of students believed the bots' advise was valuable and that they could deliver much more feedback than teaching professionals. In another research, Wessel et al. [49] addressed some of the problems with existing GitHub bots, for example, the human-bot interaction on the pull-requests can be disruptive, spam-like, and eventually be ignored by the contributors. In their paper [49], they envisioned the concept of a meta-bot which would act as a middleman between the contributors and the existing bots. Instead of handling specific tasks on pull requests, the meta-bot would provide a centralized control by integrating and orchestrating the task-oriented bots, and thus adding more value to the interaction of already existing bots.

### 2.3 Analyzing Marketplaces in Software Engineering

In recent years, app marketplaces have drawn a lot of attention among researchers within the software engineering community. In [16], Harman et al. pointed out that app marketplaces were different from traditional ones (for example, they do not have source code), and provided a framework for app marketplace analysis. Their framework consisted of extracting and refining data, extracting feature information from textual descriptions, as well as computing metrics for analysis such as correlation analysis. They also analyzed the technical, customer, and business aspects of some apps in the BlackBerry app store. Chen et al. [8] proposed ARMiner, a novel framework for app review mining, in order to extract valuable information from user reviews with minimal human efforts. After filtering out non-informative reviews using a pre-trained classifier, they employed topic modeling and a ranking scheme to automatically prioritize the informative reviews.

Al-Subaihin et al. [1] studied the app store from the developers' perspective. Through developer interviews and surveys, they tried to better understand developers' practices when making apps in order to determine the extent to which information from app stores affected developers' decision-making and to highlight open issues and challenges. They pointed out that researchers and practitioners from several software engineering sub-fields such as requirements, testing, software repository mining, etc could benefit from the findings of the survey. Nayebi et al. [36] showed that considering only app stores for review mining lacked a significant amount of feedback, rather multiple sources must be considered to get more critical and objective views of apps. For example, in their study, they found strong correlations between the number of reviews and the number of tweets for most apps. Maalej et al. [28] discussed the importance of user feedback and its relation with the software requirements and features. These feedback and the user experiences could inspire test cases, alternate flows, or even new features. Krüger et al. [24] described a preliminary analysis of the Eclipse marketplace with a view to providing a glance on open marketplaces as well as initiating further research. They proposed to mine marketplace data to address questions such as, who contributes to successful plug-ins, in order to identify leading developers and communities, leading to collaborations and new research directions. Li et al. [26] performed a performance and usability evaluation of five open-source IDE plugins that identify and report security vulnerabilities. They investigated the types of vulnerabilities these plugins could detect, quality of detection, and user-friendliness of plugin outputs. They found a mismatch between the claimed and actual coverage of the plugins, while most plugins had a high false-positive rate, and were not developer-friendly.

### 2.4 Mobile App Stores

Mojica et al. [32] presented a study where they examined whether an app rating accurately reflects how the users perceive it. Their findings indicated that the app store's current metric fails to capture the users' changing levels of satisfaction as the app evolves. Ali et al. [3] conducted a large-scale comparative study of cross-platform apps collected from the Apple and Google Play app stores. The goal was to understand the differences in how users perceived the same app distributed through different platforms. Comparing app-store attributes, such as stars, versions, and prices, and measuring the aggregated user-perceived ratings, they found many differences across the platforms. Nayebi et al. [34, 39] conducted surveys with users and developers to understand the release strategies used for mobile apps and found that an app's strategy affects its success and how it is perceived by the users. Carreño et al. [7] extracted user requirements from comments on Google Play using a sentiment-aware topic model [18]. Their method aided requirements summarisation with significantly less effort than manual identification. Khalid et al. [21] studied the user's perspective of iOS apps by analyzing app reviews. They found 12 types of user complaints including functional errors, requests for additional features, and app crashes. They also highlighted the importance of regression testing before updating apps by showing that almost 11% of the studied complaints were reported after a recent update.

## 3 Research Methodology

Our research methodology consists of three major parts. First, for **RQ1**, we used empirical protocols to retrieve information from the GitHub Marketplace as a software repository and explored and derived observations. This part mainly consists of statistical methods for descriptive analytics. Second, for **RQ2**, we employed empirical protocols for systematic mapping studies and performed a comprehensive literature analysis to extract the spectrum of software engineering research for automation tools in open source. Finally, for **RQ3**, we compared the state of the art with the state of the practice by using our findings from **RQ1** and **RQ2** and discussing the gaps and trends. In Section 3.1 we discuss our data retrieval and analysis method from the marketplace. Next, we discuss the mapping protocol in Section 3.2. The gap analysis is presented in section 3.3.

### 3.1 Process of Gathering Data from the Marketplace

To gather apps and actions from the GitHub Marketplace, we used Selenium[5] (for automated loading of the JavaScript components) along with Scrapy[6] (as the main web page crawler). Each web page loads only 1,000 Apps or actions. While this had no implication for the apps data (since there are less than a 1,000 Apps in the marketplace), this restriction meant that we could extract only a fraction of the actions data from a static page.

To avert this issue, we designed a set of knitted crawlers that not only started from the marketplace's main page but also took each action category as a scrapping root. For most categories, there were less than 1,000 Actions on the marketplace, thus we did not face any issue in gathering all the information. However, for four categories, there were more than 1,000 Actions. As a result, we used the embedded marketplace data sorting options- 'Best Match', 'Recently added', and 'Most installed/starred' (see Figure 1). Each sorting method displayed a different set of data, allowing us to store additional actions not previously found.

Apps and actions detail pages differ from each other. Also, actions have a free outline and style using GitHub tools which each developer can design. To ensure that we are gathering all the existing metadata for apps and actions, we parsed through the different HTML header layers for apps and actions. As developers provide different levels of detail about each action, we created a unified list of all the data entries provided across different actions on GitHub.

We could uniformly gather 11 Action attributes and 13 App attributes across the marketplace. Figure 2 includes details of these attributes. Some of the apps and actions had invalid URLs to their detail page, resulting in missing details for those items. However, we kept record of such items in our data collection process. Figure 4 shows the categories for apps and actions. The results of this analysis is presented in Section 4.

[5] https://www.selenium.dev/
[6] https://scrapy.org/

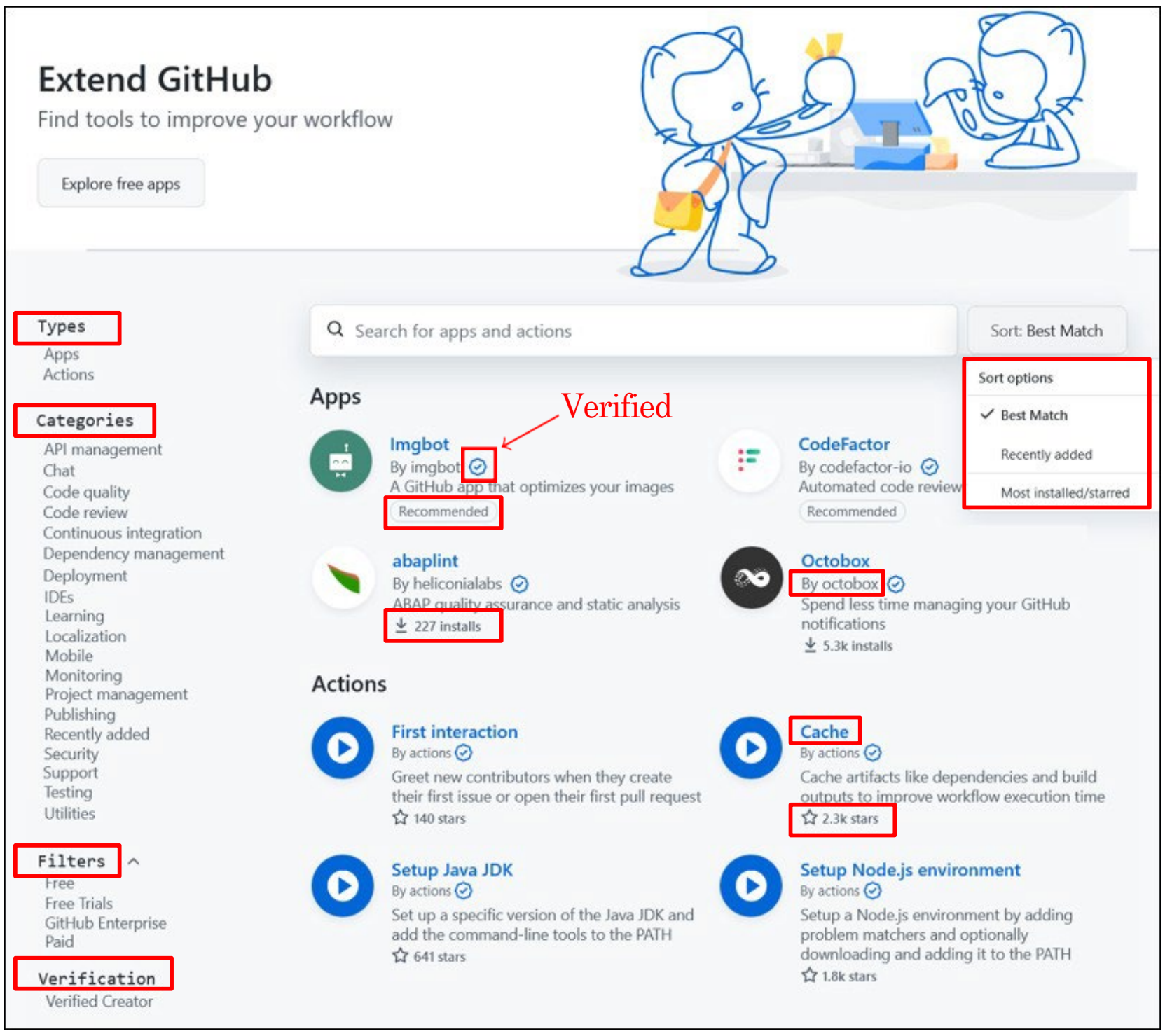

**Fig. 1** Screenshot of the GitHub marketplace with the meta data fields (Types, Categories, Filters, Verification status, Number of Installs, Number of Stars, and Developers' name) annotated.

## 3.2 Protocols Used for Systematic Mapping Study

To answer **RQ2**, we conducted a systematic mapping study with a view to comparing current trends in the open source research community with the ones in practice. To perform this systematic study, we followed the guidelines of Peterson et al. [40]. In its nature, systematic mapping is aimed at performing a general analysis of the topics and trends in the field, which perfectly serves the purpose of comparison between state of the art and state of practice we formulate in the **RQ3**. An overview of the process is shown in Figure 3. Our initial search resulted in 11,365 Research papers. By applying our exclusion criteria, we narrowed it down to 365 Papers, which we used in our final mapping. Next, we detail the process of paper selection along with the inclusion and exclusion process.

### 3.2.1 *Source Selection*

Our objective was to map the state of the art software engineering to the current categories of apps and actions in GitHub Marketplace. As such, we chose broad search terms to gather all the papers focusing on open source software engineering. Our search query was as following:

*("Software Engineering" OR "SE") AND ("GitHub" OR "OSS" OR "open source" OR "open-source.")*

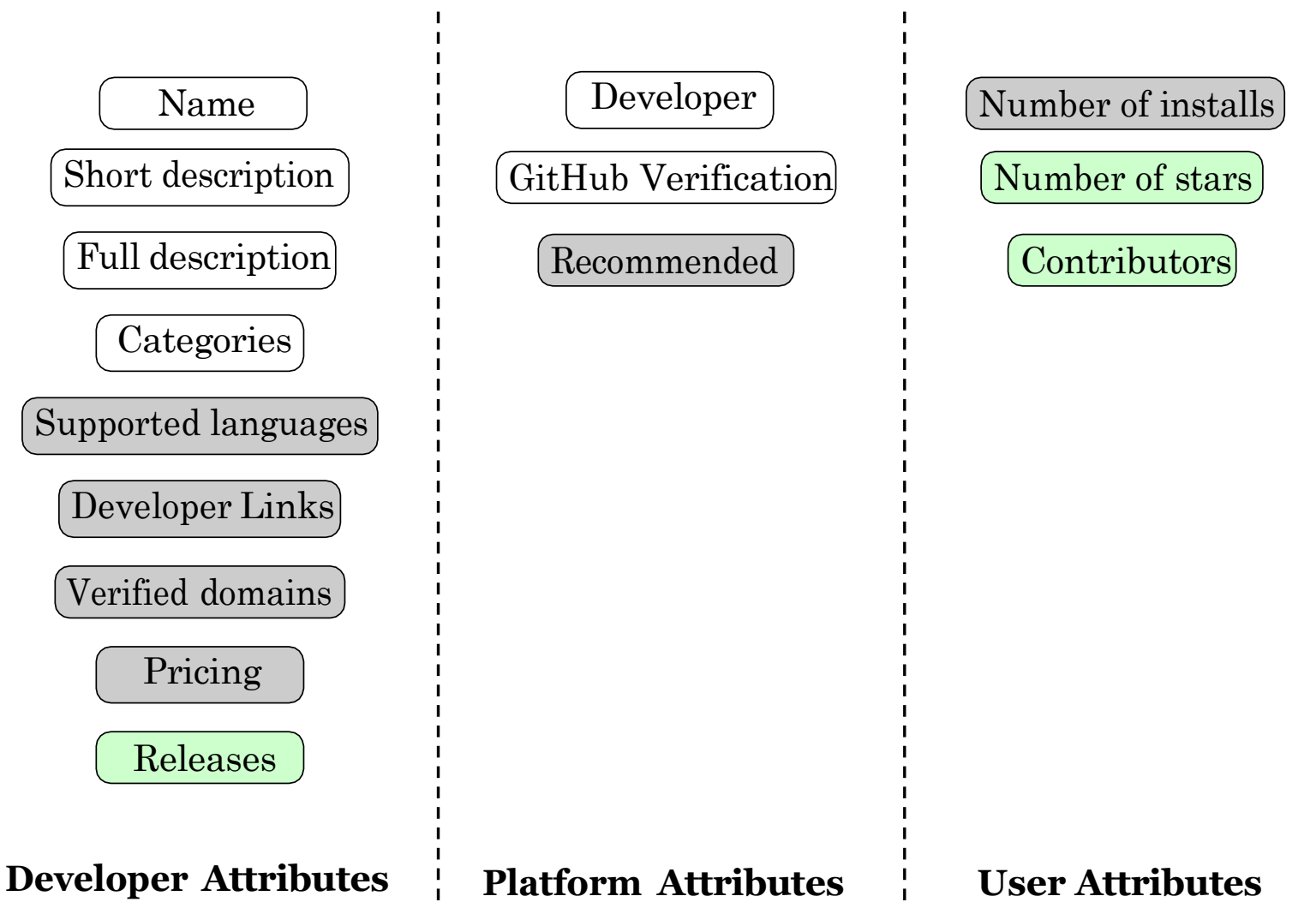

**Fig. 2** Attributes on the GitHub Marketplace for apps (gray), actions (green), or both (white). Each attribute is either defined by the developer, platform, or user.

Following previous studies, we gathered papers from five publication sources, namely ACM Digital Library, IEEE Xplore, ScienceDirect, Scopus, and Inspec. Table 1 shows the initial number of papers retrieved from each of these sources based on our search query. To ensure search quality, we applied two broad *Inclusion criteria* while performing search in these libraries:
Inclusion Criteria I: Studies must have been published between 2000 and 2021,
Inclusion Criteria II: Studies must have been published in journals or conference proceedings. Overall, our initial search resulted in 11,365 Publications.

### 3.2.2 Exclusion Criteria

To ensure quality of the publications, we used a set of heuristic exclusion criteria and narrowed down the search. We excluded the below papers:
Exclusion Criteria I: Studies not presented in English,
Exclusion Criteria II: Duplicate studies,
Exclusion Criteria III: Studies not presented in the top 20 Publication venues according to

**Table 1** The initial number of studies retrieved from five libraries using our search query with the applied inclusion criteria.

| Publication source | Num. of Studies |
|---|---|
| ACM Digital Library | 8,817 |
| IEEE Xplore | 399 |
| ScienceDirect | 1,154 |
| Scopus | 354 |
| Inspec | 619 |

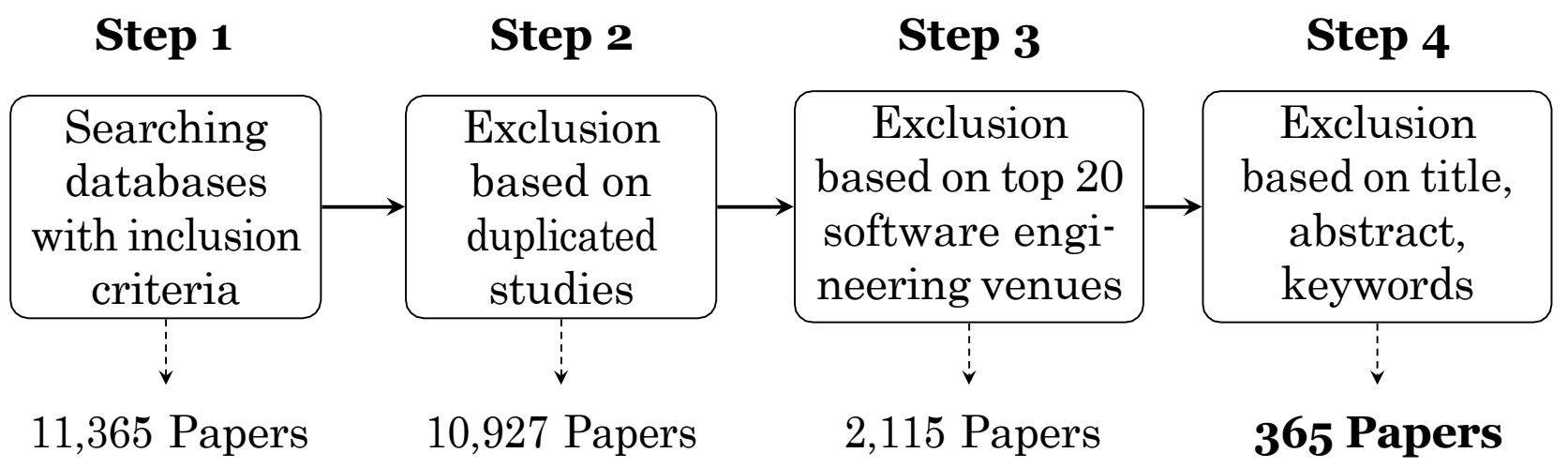

**Fig. 3** The process of our systematic mapping with the number of retrieved papers at each step.

Google Scholar's ranking[7]. The list of the venues we used as for 2021 is provided in Table 2,
Exclusion Criteria IV: Studies that do not have *software* and its variations in title, abstract, keywords and *automation* and its variations in title, abstract, and keywords.

We retrieved a total of 365 Papers after applying the exclusion criteria. Majority of the papers (10.96%) were published in 2019 followed by 10.32% published in 2017. The details of the number of publications are provided in Table 2.

**Table 2** Selected publication venues from Google Scholar as used for exclusion Criteria III and the final number of papers studied from each library.

| Rank | Publication | #Papers |
|---|---|---|
| 1 | ACM/IEEE Inter. Conference on Software Engineering | 0 |
| 2 | Journal of Systems and Software | 106 |
| 3 | IEEE Transactions on Software Engineering | 101 |
| 4 | Information and Software Technology | 111 |
| 5 | ACM SIGSOFT Inter. Symposium on Foundations of Software Engineering | 0 |
| 6 | Empirical Software Engineering | 28 |
| 7 | IEEE Software | 12 |
| 8 | ACM SIGPLAN Conference on Programming Language Design and Implementation (PLDI) | 0 |
| 9 | Mining Software Repositories (MSR) | 0 |
| 10 | IEEE/ACM Inter. Conference on Automated Software Engineering (ASE) | 0 |
| 11 | ACM SIGPLAN-SIGACT Symposium on Principles of Programming Languages (POPL) | 0 |
| 12 | Inter. Conference on Software Analysis, Evolution, and Reengineering (SANER) | 0 |
| 13 | Proceedings of the ACM on Programming Languages | 0 |
| 14 | Software & Systems Modeling | 3 |
| 15 | Inter. Symposium on Software Testing and Analysis | 0 |
| 16 | IEEE Inter. Conference on Software Maintenance and Evolution | 0 |
| 17 | Software: Practice and Experience | 4 |
| 18 | Conference on Tools and Algorithms for the Construction and Analysis of Systems (TACAS) | 0 |
| 19 | IEEE Inter. Requirements Engineering Conference | 0 |
| 20 | ACM SIGPLAN Symposium on Principles & Practice of Parallel Programming (PPOPP) | 0 |

### *3.2.3 Data Extraction and Synthesis*

For the classification scheme, we used the GitHub Marketplace categories. In the marketplace, the product owner defined one or more categories for each app and action. Categories are a common practice in marketplaces that offer end users a convenient way to find products and services and possibly compare and choose among different alternatives. GitHub apps and actions can be classified in one or more categories.

[7] https://scholar.google.ca/citations?view_op=top_venues&hl=en&vq=eng_softwaresystems

We followed a protocol for the manual categorization of the papers. First two co-authors of the paper each carefully reviewed 365 Paper abstracts and categorized them up to three categories. The classification results were compared, and a total of 79 Disagreements were found in at least one of the categories. The two authors discussed these conflicts and could converge their opinion for 14 of these papers. In the case that no agreement could be made, the third co-author was involved. The third co-author was provided with the list of 65 Papers and independently categorized them. Then, the categorized results were reviewed by the authors and the list was finalized. As the result of this process, 53 Papers were identified by all the authors as irrelevant to the scope of this study and could not be mapped into any marketplace category. Of the remaining 312 Papers, the authors identified 20 Papers as systematic literature reviews and systematic mapping studies and excluded them from the mapping study. Hence, the answer to **RQ2** as detailed in Section 5 is based on 292 Papers.

### 3.3 Gap Analysis

We performed quantitative and qualitative comparison to diagnose the gap between research and practice. First, we compared the population of researchers and practitioners publishing papers or tools related to a particular category. We calculated the number of distinct paper authors and distinct app and action developers in our data of **RQ1** and **RQ2**. Although on GitHub we identified developers with their unique user IDs, we found several different formats for author names from the studies in our mapping study (RQ2). Hence, we performed substring matching to ensure that we identify these different formats and avoid counting duplicate authors. On the other hand, there can be multiple authors having the same name. In order to distinguish between authors with identical names, we performed manual inspection of their research profile on the internet, compared their affiliations, co-authors, and then made a decision. For each category in the marketplace, we then compared the number of authors to the number of developers. We highlighted categories where a higher proportion of researchers are working compared to developers and vice versa. We also identified the categories with the least difference between the proportion of researchers and developers.

Second, we counted the number of contributions (i.e the apps and actions in the marketplace (**RQ1**) and the number of papers in literature (**RQ2**)) to discuss the extent of topic coverage. We compared the absolute number and the percentage (for example, proportion of the number of papers in each category to overall number of papers).

Third, we compared the popularity of studies with the tools in the marketplace. As for the popularity in practice, we used the average number of installs (for apps), and the average number of stars (for actions) per each category as the proxy. For apps, we got the average number of installs per category by taking the total number of installs for all apps in a category and dividing it by the total number of apps in that category. We calculated the average number of stars per category in a similar manner. For the academic popularity, we calculated the average number of citations *per year* across all the papers. We used Google Scholar's API to gather the number of citations for each paper. In order to get the average number of citations per year, we divided the number of citations of a paper by the number of years from the time they were published. We used this number instead of the total number of citations so that the papers that were published earlier did not get an unfair advantage. We then calculated the average number of citations per category in the same way as we calculated the other two metrics for apps and actions. We ranked each category based on these popularity proxies.

Fourth, we compared co-occurrence and overlap between categories among the marketplace tools (**RQ1**) and the mapped studies (**RQ2**). Developers can categorise their apps and actions in more than one category within the Marketplace. Similarly, within our mapping study we categorized the papers in up to three categories. Often, new technologies and innovations appear from the synergy between topics. Hence, we compared the perspective of researchers and practitioners on the relevance and connection between different categories. Also, we identified the most active and inactive intersection between the categories.

Fifth, we compared the content and description of marketplace tools with the state of the art studies. We used the short descriptions and full descriptions of apps and actions to extract the key topics in the marketplace (**RQ1**). For literature, we extract these topics from the paper title and abstracts in **RQ2**.

## 4 RQ1: Observations from Mining GitHub Marketplace

As the first study on analyzing the GitHub Marketplace, our first research question (**RQ1**) is focused on reporting observations from this platform. To this end, we performed an explorative study on all the available data attributes and performed group comparisons whenever possible. In contrast to the known app stores for the software engineering research community, the GitHub Marketplace provides moderate degree of customization for the developers. This results in non-unified information across apps and actions within the marketplace. For instance, only 5.01% of apps provide the information of their customers. In this study, we focused on the information that is available for all the apps and actions.

We scraped data for 440 Apps and 7,878 Actions. However, we could not retrieve the details of one app and 32 Actions due to HTTP status code errors. As a result, the following results are based on analysis of 439 Apps and 7,846 Actions. We first compare differences between apps and actions. Then, we discuss the results of our observations based on the characteristics of the information (attributes) provided by the platform, developers, and users (see Figure 2).

### 4.1 Apps Versus Actions in GitHub Marketplace

GitHub apps are a particular type of GitHub action that are hosted and run by GitHub. Apps are mostly used in the case that:

- Accessing the team's workflow requires further permissions to the repository,
- Repository owner would need to host and run the automation, or
- There is a need for a solid architecture of the hosted data.

GitHub actions are dependent on runners to execute GitHub action workflows. One can host their own runner and customize the environment for GitHub action workflows (self-hosted runners) or use GitHub virtual machines to run workflows which contains tools, packages, and settings available for GitHub actions (GitHub-hosted runners). In comparison to GitHub hosted runners, the self-hosted runners offer more control of hardware, operating system, and tools[8,9].

Different steps are required for adding apps and actions to the GitHub Marketplace. To be successfully listed on the marketplace, apps must meet *user experience requirements*, *branding*

[8] https://docs.github.com/en/actions/using-github-hosted-runners/about-github-hosted-runners

[9] https://docs.github.com/en/actions/hosting-your-own-runners/about-self-hosted-runners

*requirements*, in addition to *billing requirements* for paid apps. The *user experience requirements* state that apps must not sway users away from GitHub, must integrate with the platform beyond authentication, and must be publicly available on the marketplace. Also, listings must include valid contact information for the publisher, who should set up webhook events to get notified about GitHub plan changes or cancellations through the Marketplace API.

Moreover, apps must have a relevant description and must specify a pricing plan. According to the *branding requirements*, apps must have a logo, feature card, screenshot images, and a well written description which are aligned with the recommendations provided by GitHub. Paid apps also have additional requirements. To publish a paid plan for an app (or an app that offers at least one paid plan) on the GitHub Marketplace, the app must be owned by an organization and the app developer must have owner permissions in the organization. Furthermore, the organization must be verified. At least 100 Installations and 200 Users are required for GitHub apps and OAuth apps respectively to be listed as a paid app. Under the *billing requirements*, all paid apps need to handle GitHub Marketplace purchase events to manage new purchases, upgrades, downgrades, cancellations, and free trials.

To publish an action, a developer first needs to create the action in their repository. They need to ensure that the repository only includes the metadata file, code, and files necessary for the action. This allows the developer to tag, release, and package the code in a single unit. The action must be in a public repository and the repository must contain a single action. The action's metadata file should have a unique name and be located in the root directory of the repository. The name cannot match an existing GitHub Marketplace category as GitHub reserves the names of GitHub features. If these requirements are met, the developer can add the action to the GitHub Marketplace by tagging it as a new release and publishing it immediately on the GitHub Marketplace.

*GitHub Marketplace offers apps and actions that provide various attributes of developers as users. While in the mobile app stores, the user is often a general public, in this marketplace, the users of the offered products are software developers as well.*

## 4.2 Developers Attributes

As developers publish their products (apps and actions) on the marketplace, they define several attributes for their products. In this section, we discuss these developer attributes.

### *4.2.1 Name*

Names are the unique identifier for products (apps and actions) on the marketplace. The name is limited to 255 Characters. Products cannot have the same name as an existing GitHub account unless it is the developer's own user or organization name. For actions, the name should not match an existing marketplace category nor can it have the same name as a published action.

### *4.2.2 Categories*

According to GitHub documentation, each app and action in the marketplace can be listed in one or two (primary and secondary) categories by its developer. The primary category should best represent the main functionality of the product, while the optional secondary category should also fit the product. While this is true for actions, we have observed that apps sometimes have more than two categories. 39 Apps have three categories and two apps have four categories. Currently 32 Categories exist in the marketplace where two of them are only exclusive to apps ('Content attachments API', and 'GitHub Created') and two are exclusive to actions ('Community', and 'Game CI'). The detailed list of all these categories is provided in Table 3, which shows that 87.5% of the categories are mutual between apps and actions.

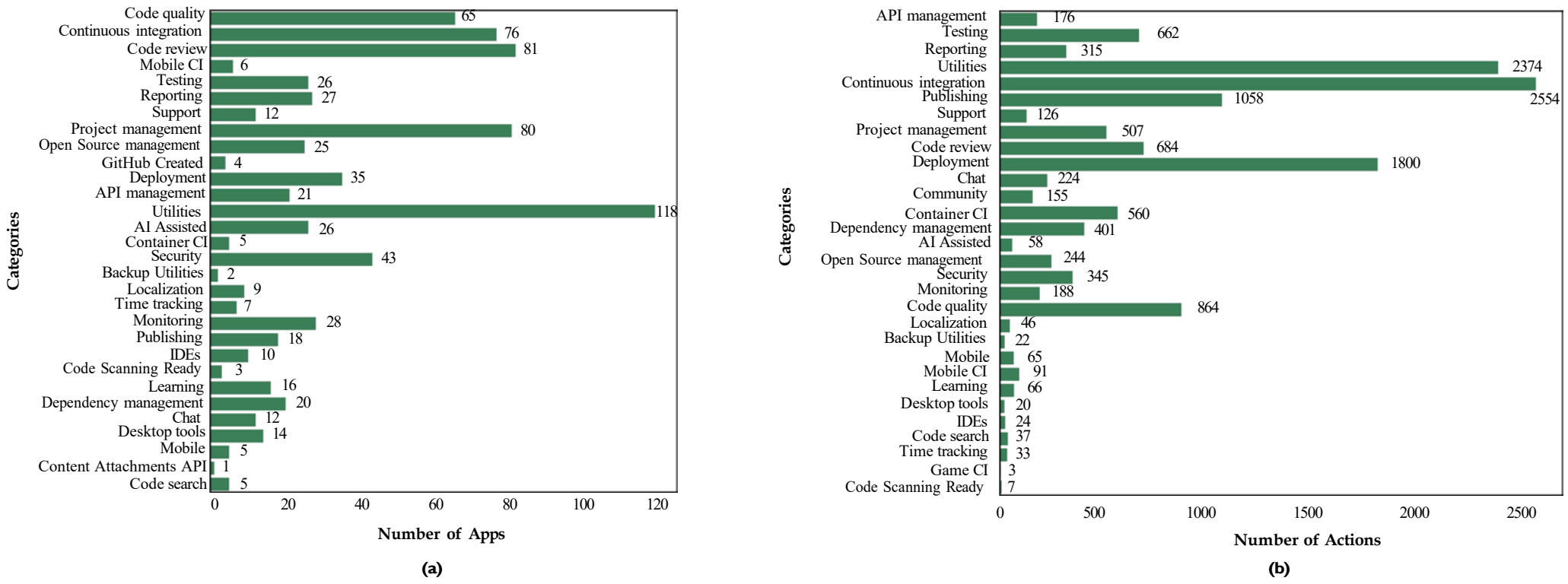

**Fig. 4** Each app or action is listed in one or more categories by its developer. (a) shows the categories and frequencies for apps, and (b) shows the categories and frequencies for actions.

Figure 4 shows the number of apps and the number of actions in each category. We used these categories to further answer **RQ2** and systematically map literature on open source software and automation into these categories. For apps, categories 'Utilities' (118 Apps), 'Code Review' (81), and 'Project Management' (80) has the highest population. While for actions, 'Continuous Integration' (2,554 Actions), 'Utilities' (2,374), and 'Deployment' (1,800) are the categories with the most population.

As a means to understand the similarity between the automated tools provided on the marketplace, we were interested in examining how often two categories appeared together and how many apps or actions the categories involved. Figure 5 demonstrates the frequency of categories chosen together for apps in Figure 5-(a) and actions in Figure 5-(b). For apps, the biggest intersection belongs to 'Code Quality' and 'Code Review' categories with 31 mutual apps. This is followed by 'Utilities' and 'Project management' with 26 Apps. 'Utilities' pairs with 23 out of 29 other app categories, followed by 'Project Management' (with 19 other categories), 'Continuous integration' (with 19 other categories), 'Code review' (with 17 other categories), and 'Security' (with 15 other categories). On the other hand, 'Code Scanning Ready' and 'Content Attachments API' each pair with only one other app category, 'Security' and 'Project management', respectively.

**Table 3** Categories of apps and actions on the GitHub Marketplace. *We will refer to 'API management and Checking' as 'API management' in the subsequent discussions

| ID | Type | Category | ID | Type | Category |
|---|---|---|---|---|---|
| C1 | Both | API management and Checking* | C17 | Both | Security |
| C2 | Both | Testing | C18 | Both | Monitoring |
| C3 | Both | Utilities | C19 | Both | Code quality |
| C4 | Both | Reporting | C20 | Both | Localization |
| C5 | Both | Continuous integration | C21 | Both | Desktop tools |
| C6 | Both | Publishing | C22 | Both | Mobile |
| C7 | Both | Support | C23 | Both | IDEs |
| C8 | Both | Project management | C24 | Both | Mobile CI |
| C9 | Both | Code review | C25 | Both | Code search |
| C10 | Both | Deployment | C26 | Both | Code Scanning Ready |
| C11 | Both | Chat | C27 | Both | Learning |
| C12 | Action | Community | C28 | Both | Time tracking |
| C13 | Both | Container CI | C29 | Action | Game CI |
| C14 | Both | Dependency management | C30 | Both | Backup Utilities |
| C15 | Both | AI Assisted | C31 | App | Content Attachments API |
| C16 | Both | Open Source management | C32 | App | GitHub Created |

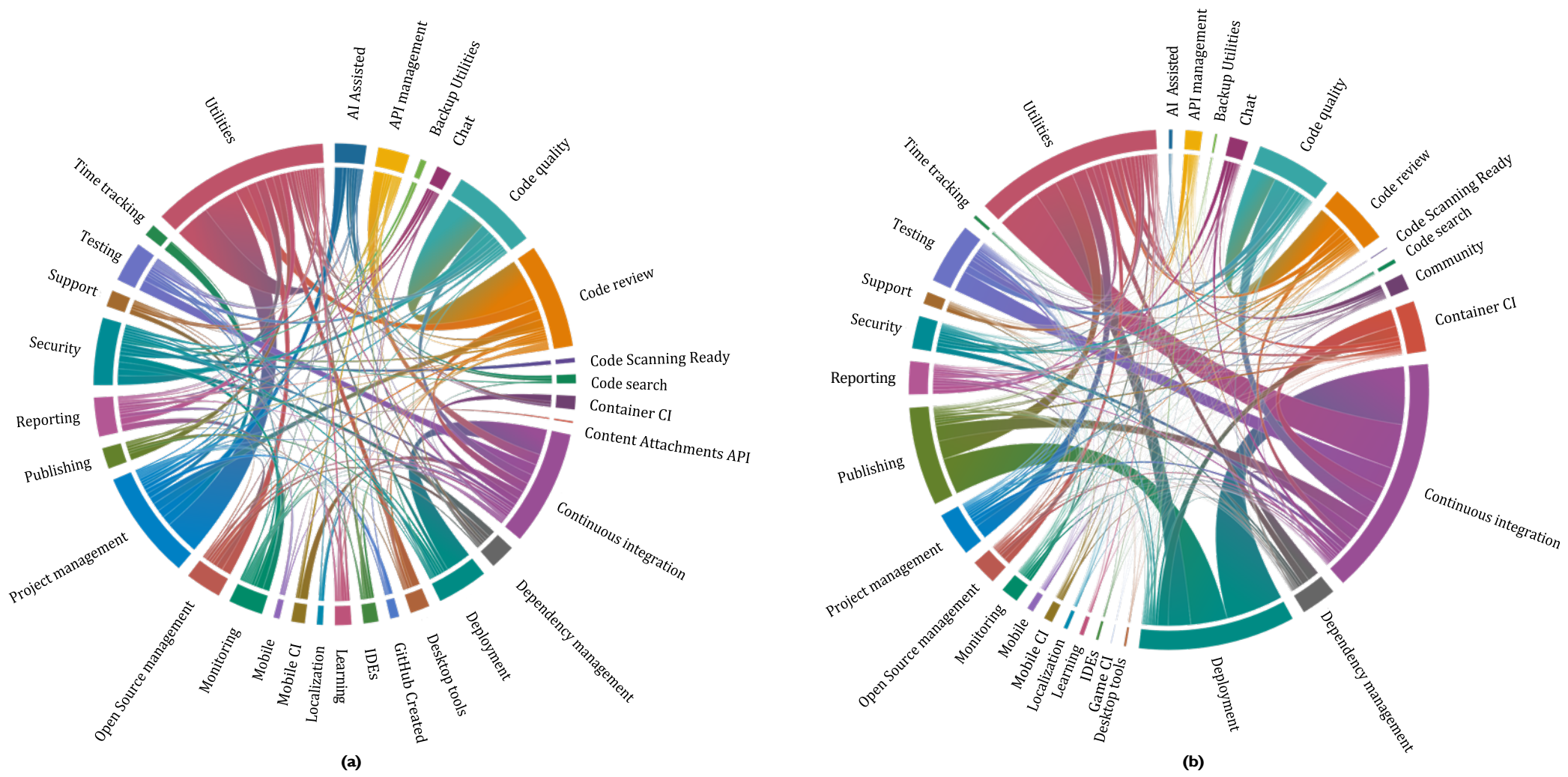

**Fig. 5** Frequency of categories appearing together for **(a)** apps and **(b)** actions on the marketplace.

For actions (Figure 5-(b)), 'Continuous integration' and 'Deployment' appear together for 603 Actions followed by 'Continuous integration' and 'Utilities' with 466 Actions. The category 'Utilities' pairs with all other (28 out of 29 Categories) action categories except 'Code Scanning Ready'. It also appears that 'Continuous integration' (with 27 other categories), 'Code quality' (with 25 other categories), 'Publishing' (with 25 other categories), and 'Testing' (with 25 other categories) has high number of intersections and synergy. 'Game CI' (with three other categories), 'Code Scanning Ready' (with four other categories) and 'IDEs' (with seven other categories) pair with the least number of other categories.

*Developers classify their provided tools in up to two of the the 32 Categories provided in the marketplace. There are two categories exclusive for apps and two are exclusive to actions. 'Utilities' include most number of apps while 'Continuous Integration' has the most number of actions. We use these categories for mapping studies in **RQ2**.*

#### 4.2.3 Short description

GitHub allows product owners to add brief descriptions for their products in the marketplace to explain the product's main functionality. GitHub suggests keeping short descriptions at 40-80

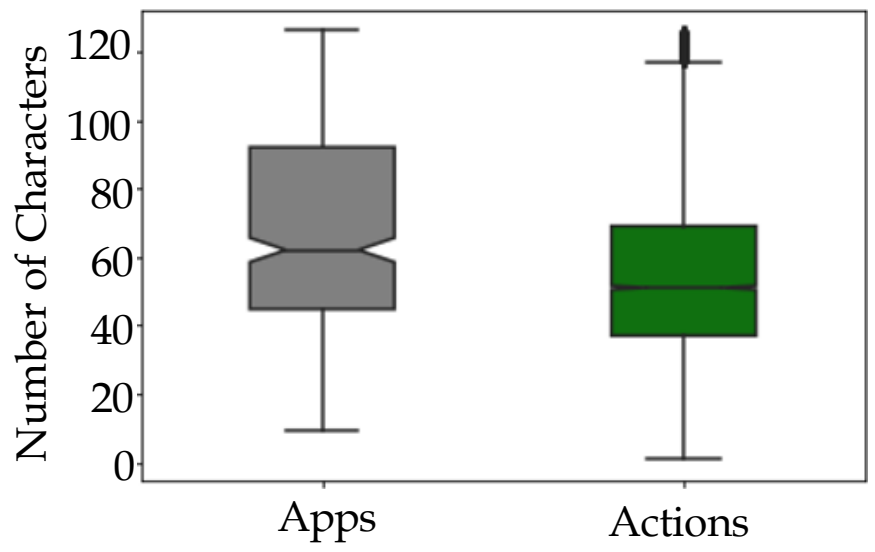

**Fig. 6** Number of characters in short descriptions for apps and actions on the GitHub Marketplace

Characters. GitHub also recommends that short descriptions do not include complete sentences or more than one sentence[10].

Figure 6 shows that on average apps have longer short descriptions compared to actions. On average, apps have short descriptions of 68.72 Characters and 10.54 Words and actions have short descriptions of 55.13 Characters and 8.97 Words. 18.64% of the apps (82 out of 440) have short descriptions of less than 40 Characters while 33.64% of the apps (148 out of 440) have short descriptions that exceed 80 Characters. This means that 52.28% Apps on the marketplace have a short description length that falls outside the GitHub recommended range of 40-80 characters. For actions, 28.90% and 15.65% have a short description length of below 40 Characters and above 80 Characters, respectively. There is a very weak correlation between short description lengths and number of stars for actions ($correlation = 0.04$) and number of installs for apps ($correlation = -0.01$).

We also were interested to see if specific (key)words and bigrams in the product description contributed to higher number of installs (apps) or stars (actions). With this goal, we vectorized the app short description texts using TfidfVectorizer with sublinear `tf` scaling, which considers the logarithm of the term frequency, and ngram_range set to (1, 2), meaning that we want unigrams and bigrams. We then fit a linear regression model with these vectors against the number of installs. The regression model determines the optimal weights to best fit the data. Observations that have relatively larger weights have more influence in the analysis than observations that have smaller weights. Finally, we used the *Eli5* [11] python package to inspect model parameters and understand how the model works globally. Specifically, we used the show_weights function from the package which takes our model and vectorizer as inputs and returns an explanation of model parameters (weights). In other words, this function gives us the words that contribute most to the number of installs, either positively or negatively. Words such as 'learn', 'connect', 'deploy' have the largest positive weights while words such as 'machine', 'help', 'open source' negatively affect the number of installs. We also ran the same experiment with action descriptions. Table 4 lists the top words and bigrams that most affect the number of installs and number of stars for apps and actions, respectively.

#### *4.2.4 Full description*

Each product has a dedicated page and URL on the marketplace which includes an elaborate description of the app or action. For actions, GitHub does not provide guidelines regarding the format or length of this description. However, for apps, this full description should consist of two parts: a required 'Introductory description', and an optional 'Detailed description'[12].

[10] https://docs.github.com/en/developers/github-marketplace/listing-an-app-on-github-marketplace/writing-a-listing-description-for-your-app#very-short-description

[11] https://eli5.readthedocs.io/en/latest/index.html

**Table 4** Words and bigrams with the highest positive and negative weights impacting number of installs (apps) and number of stars (actions). Green shows positive and red shows negative weights while darker shade indicates greater amounts.

| Apps | | Actions | |
|---|---|---|---|
| **Word/Bigram** | **Weight** | **Word/Bigram** | **Weight** |
| learn | +392817.486 | linters | +2639.525 |
| connect | +72359.028 | validate yml | +1527.648 |
| deploy | +58359.672 | detect bug | +996.803 |
| repo | +35675.965 | git history | +900.346 |
| speed | -33828.643 | process project | +813.082 |
| code | -41012.257 | option | +682.447 |
| apps | -45310.336 | smell program | +678.411 |
| open source | -50067.428 | bash | +624.838 |
| help | -67253.429 | ssh rsync | -620.961 |
| machine | -131001.443 | jupyter notebook | -2071.856 |

The 'Introductory description' is displayed at the top of the app's landing page. Although developers are free to use more characters, GitHub advises writing a one or two sentence high-level summary between 150-250 Characters in the 'Introductory description' field. Developers can provide further details in the 'Detailed description' section, which consists of three to five highlights about their products' value (or defined by GitHub as the 'value propositions'), each described in no more than two sentences. A value proposition is a short statement that summarizes why a GitHub user should use a particular app. GitHub recommends that developers include a title and a paragraph of description for each value proposition and avoid complete sentences or more than one sentence in value proposition titles. For detailed descriptions, developers can use up to 1,000 Characters. Developers can also add up to five screenshots of their app to the landing page. After they've been uploaded, developers can rearrange the screenshots and add captions to each. GitHub also provides guidelines for screenshots[13].

To understand the features offered by the tools in each category, we mined the long descriptions of all the tools in each category. Similar to the work of Al-Subaihin et al. [2] on mining features from mobile app descriptions, we refer to a feature as a claimed functionality mined from the product description. We pre-processed the text by first removing URLs, emojis, `html tags`, and punctuation from the descriptions. We also built a thesaurus for mapping abbreviation and replaced each abbreviation with a full text. For this, we manually built a list of abbreviations by scanning through the most frequent 500 Words in long descriptions. From this preprocessed data, we extracted the top words and bigrams in each category, the latter often resembling product features. We compared the extracted bigrams to further investigate the overlap between the categories. Table 5 shows the top features from each category and the frequency in which they appear in product descriptions. The most frequent feature of the products in the GitHub Marketplace is `issue management`. Categories 'Project management' (127 Mentions of this feature), 'Open Source management' (54), 'Utilities' (53), 'AI Assisted' (21), and 'Support' (10) all have products that offer this feature to some extent such as open or close issue, label issue, find issue, or comment on issue using bot. Among these, labeling issue is the most prevalent feature among products. The feature `unit test` also appears frequently in products from the categories 'Testing' (103), 'Reporting' (50), 'AI Assisted' (12), 'Mobile' (9), 'Continuous integration' (6), 'Code search' (5), and 'Deployment' (3). Apps and actions in the marketplace which facilitate static analysis mainly fall in 'Code quality' (72), 'Security' (21), 'Code review' (3), and 'Continuous integration' (3). The categories 'Dependency management' (17), 'AI Assisted' (11), 'Utilities' (11), and 'Open Source management' (6) are also similar in that they all have products that help with managing dependency. These products mainly aid software developers in updating dependencies, resolving dependency conflicts, creating dependency graphs and documentations. A variety of products in overlapping categories also help developers with code coverage, API documentation, automating deployment, detecting and fixing vulnerabilities in software, and building and maintaining Docker images.

[12] https://docs.github.com/en/developers/github-marketplace/listing-an-app-on-github-marketplace/writing-a-listing-description-for-your-app#listing-details

[13] https://docs.github.com/en/developers/github-marketplace/listing-an-app-on-github-marketplace/writing-a-listing-description-for-your-app#product-screenshots

**Table 5** Top features in each marketplace category (extracted from the long description of products). The numbers in parenthesis stand for the frequencies of each feature appearing in the product descriptions.

| Category | Top features |
|---|---|
| AI Assisted | label issue (21), unit test (12), dependency conflict (11), search code (4), code enhance (3) |
| API management | api specification (15), api documentation (13), swagger ui (13), performance test (5), schema change (3) |
| Backup Utilities | backup repository (5), data retention (4), daily backup (3), run workflow (3), cloud storage (2) |
| Chat | send message (110), slack notification (76), custom message (34), discord webhook (30), slack bot (29) |
| Code quality | code coverage (113), run test (82), static analysis (72), code review (27), automate code (5) |
| Code review | linting process (64), code coverage (59), automatically merge (6), test coverage (5), automate code (5) |
| Code scanning ready | infrastructure code (4), code scan (4), check vulnerability (4), automatically merge (2), automate deployment (2) |
| Code search | issue create (7), unit test (5), code enhance (3), best practice (3), search code (3) |
| Community | api key (25), push branch (20), create issue (17), open source (17), issue comment (14) |
| Container CI | container registry (126), run docker (82), build image (74), push image (61), machine learning (3) |
| Content Attachments API | real time (3), account login (2), embed issue (2), team collaboration (2), collaboration leanboard (2) |
| Continuous integration | docker image (347), run test (320), run workflow (206), docker container (160), unit test (36) |
| Dependency management | run test (54), project documentation (31), dependency conflict (11), fix vulnerability (7), update dependency (6) |
| Deployment | docker image (226), run build (136), cloud deploy (4), easy build (3), automatic deployment (3) |
| Desktop tools | cake script (12), UI test (5), desktop app (5), quick filter (5), issue tracker (2) |
| Game CI | resource pack (11), optimize resource (5), build renpy (2), generate zip (2), file distribution (2) |
| GitHub Created | jira issue (4), friendly bot (3), learn skill (3), post comment (2), connect jira (2) |
| IDEs | code editor (4), smart IDE (3), code deploy (2), app inspection (2), sass solution (2) |
| Learning | update readme (8), execute workflow (7), smart IDE (3), learn skill (3), share knowledge (3) |
| Localization | check spelling (5), generate translation (4), machine translation (4), google translate (4), continuous localization (2) |
| Mobile | run test (26), android emulator (9), unit test (9), android app (8), deploy apps (3) |
| Mobile CI | run android (14), android emulator (8), android CI (7), upload artifact (6), deploy apps (3) |
| Monitoring | send notification (19), run lighthouse (16), html report (14), graphql inspector (5), crash reporting (3) |
| Open Source management | create workflow (31), close issue (25), label issue (29), post comment (5), merge pull-request (3) |
| Project management | create issue (68), label issue (59), create release (45), kanban board (9), track progress (7) |
| Publishing | create release (266), release note (178), docker image (132), build deploy (102), release tag (94) |
| Reporting | send message (52), unit test (50), slack notification (46), test report (33), see dashboard (2) |
| Security | security scan (53), scan repositories (51), security analysis (36), fix vulnerability (25), static analysis (21) |
| Support | support request (14), development support (13), label issue (10), powerful analytics (3), customer service (3) |
| Testing | unit test (103), test playbook (88), code coverage (66), test report (47), test automation (6) |
| Time tracking | todoist api (10), update readme (9), todo list (8), task organization (2), productivity growth (2) |
| Utilities | project documentation (159), docker image (128), create workflow (108), dependency conflict (11), pull-request review (8) |

*Developers provide long and short descriptions along with their tools on the marketplace. Analysis of these texts for extracting software features shows a significant overlap between functionalities for different tool categories.*

#### *4.2.5 Apps' Supported languages*

If an app only works with specific programming languages, its developer can select up to ten programming languages the app supports. These languages are displayed on the app's listing page. Filling this field is optional for app developers and is not applicable for actions. We identified 80 Distinct programming languages that are supported by apps on the marketplace. However, 234 Apps (53.30%) did not document their supported languages. Among the rest, JavaScript with 71.22% (146 out of 205 apps), Python with 59.51% (122 out of 205), Java with 57.56% (118 out of 205), Ruby with 46.83% (96 out of 205), and Go with 42.93% (88 out of 205) are the top five languages supported by the apps.

Some apps support more than one programming language. The chord diagram of Figure 7 demonstrates the co-occurrence of these languages among the apps. Most frequently, Javascript and Python appear together in 107 Apps followed by Java and Javascript which were jointly supported in 106 Apps. 77 out of the 80 distinct languages identified on the marketplace are supported by apps that support at least one other language. ABAP, LookML, and OCaml are the only languages not associated with any other languages. Javascript pairs with 60 out of 79 other languages, followed by Python (pairing with 58 other languages), PHP (with 55), Java (with 52), and Go and Ruby (with 51). On the other hand, Dhall, reStructuredText, and RMarkdown each pair with only one other language (Markdown) and each pair of languages is supported by only one app[14].

#### *4.2.6 App Developer Links*

The app listing page on the marketplace includes required and optional URLs. The two required URLs are the Customer support URL and Privacy policy URL. According to GitHub documentation[15], the privacy URL should display the app's privacy policy while the support URL should redirect clients to a web page for getting technical help, product information, or account information.

The optional URLs include company URL, status URL, and documentation URL. The company URL is a link to the company's website while the status URL is a link to a web page that shows the app's status. Current and historical incident reports, web application uptime status, and scheduled maintenance can all be seen on status pages. Finally, the documentation URL leads to documentation that demonstrates how to use the app.

[14] These are apps for a real-time collaborative Markdown editor

[15] https://docs.github.com/en/developers/github-marketplace/listing-an-app-on-github-marketplace/writing-a-listing-description-for-your-app#listing-urls

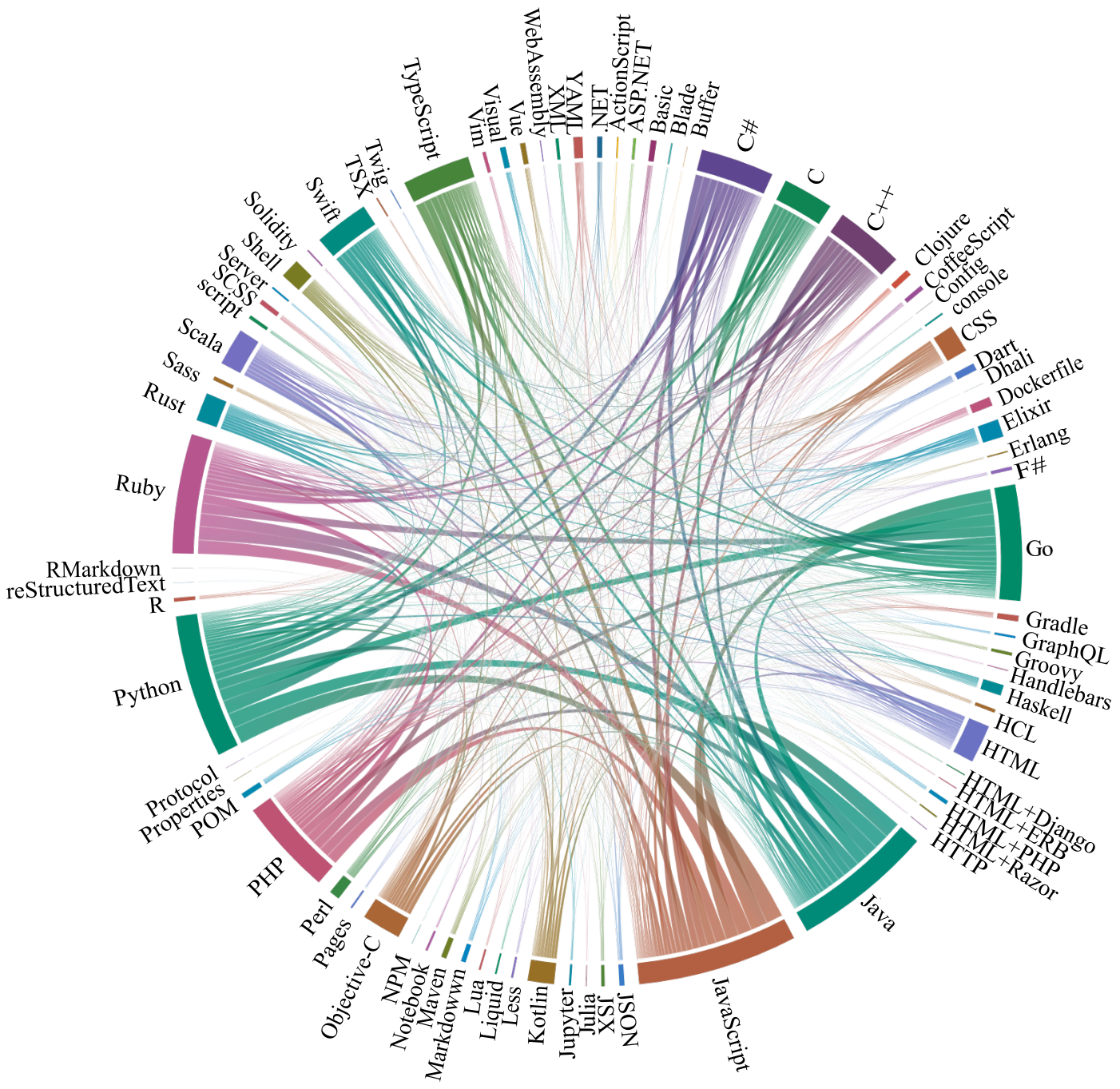

**Fig. 7** Frequency of languages appearing together for apps

#### *4.2.7 App Verified domains*

In Section 4.3.2, we mentioned that before app developers can offer paid plans for their apps or include a marketplace badge in their app listing, they are required to complete the publisher's verification process for their organization. As part of this process, the publisher must ensure that their organization has verified ownership of their domain. 43.51% (191 out of 439) Apps on the marketplace have verified domains.

#### *4.2.8 Apps' Pricing*

GitHub actions are free for both GitHub hosted runners and self-hosted runners (see Section 4.1). For instance, users can choose to create a custom hardware configuration with more processing power in order to run larger jobs using a self-hosted runner. For self-hosted runners, each GitHub account receives a certain amount of free minutes and storage. For example, *GitHub Free* has 500 MB of storage and 2,000 Minutes/Month, compared to 1 GB storage and 3,000 Minutes/Month for *GitHub Pro*[16]. Customers are billed for additional usage of GitHub actions beyond the storage or minutes included in their account.

[16] https://docs.github.com/en/billing/managing-billing-for-github-actions/about-billing-for-github-actions#included-storage-and-minutes

For apps, GitHub Marketplace pricing plans can be *free*, *flat rate*, or *per-unit*. *Free plans* are completely free for users and developers are encouraged (not enforced) to offer free plans as a way to promote open source services. There are two types of paid pricing plans- *flat rate* and *per-unit*. *Flat rate* pricing plans charge a fixed fee on a monthly and yearly basis. On the other hand, *per-unit* pricing plans charge a fixed fee on either a monthly or yearly basis for a unit that the publisher specifies such as a user, seat, or per developer. Up to 10 Pricing plans can be offered in the marketplace listing of each app. These pricing plans allow publishers to provide their app with different levels of service or resources. If an app has multiple plan options, the publisher can set up corresponding pricing plans. For example, if the app has two plan options, an open source plan and a pro plan, the publisher can set up a free pricing plan for the open source plan and a flat pricing plan for the pro plan. Different publishers choose different units or combination of units for their pricing plans. This variety of possible monetization plans along with the freedom to choose the name for the offered plan creates a very inhomogeneous monetization data for repository mining.

Currently, there are 223 different pricing plan names on the marketplace, 51 of which are variations of *Free*, for example, Free, Open source, Basic, Default, Starter, Hobby, Just enjoy it, etc. On the marketplace, 96.81% (425) Apps have a free tier compared to 19.36% (85) Apps with at least one paid plan. 80.64% (354) Apps have only the free plan in contrast to 3.19% (14) Apps that have only paid plans. Finally, 16.17% (71) Apps have both free and paid plans. The publisher may also want to provide free 14-day trials to customers. Whether the free trial is for the flat-rate or per-unit pricing plan can be specified while setting up the pricing plans for the app on the marketplace. Customers can purchase apps without leaving GitHub by using GitHub's billing API to pay for the service with the payment method that is already attached to their GitHub account.

*Most apps have a free tier while 3.19% Apps offer only paid plans. There is no uniform unit for apps' pricing plans on the marketplace which results into heterogeneous data for monetization analysis.*

*4.2.9 Actions' Releases*

Actions include release information in their page and users can view and choose to use older releases if available. GitHub documentation does not mention if there is a limit to the maximum number of releases that can be displayed on the action landing page. However, we observed that the number of releases range from 1 to 10. For each value $n$ within this range, Figure 8 shows the number of actions that have $n$ releases. Majority of the actions have more than one release. 42.64% of the actions only have one release while 1.43% of the actions have nine releases. However, the Pearson correlation coefficient (r) between number of releases and number of stars is rather weak (= 0.19).

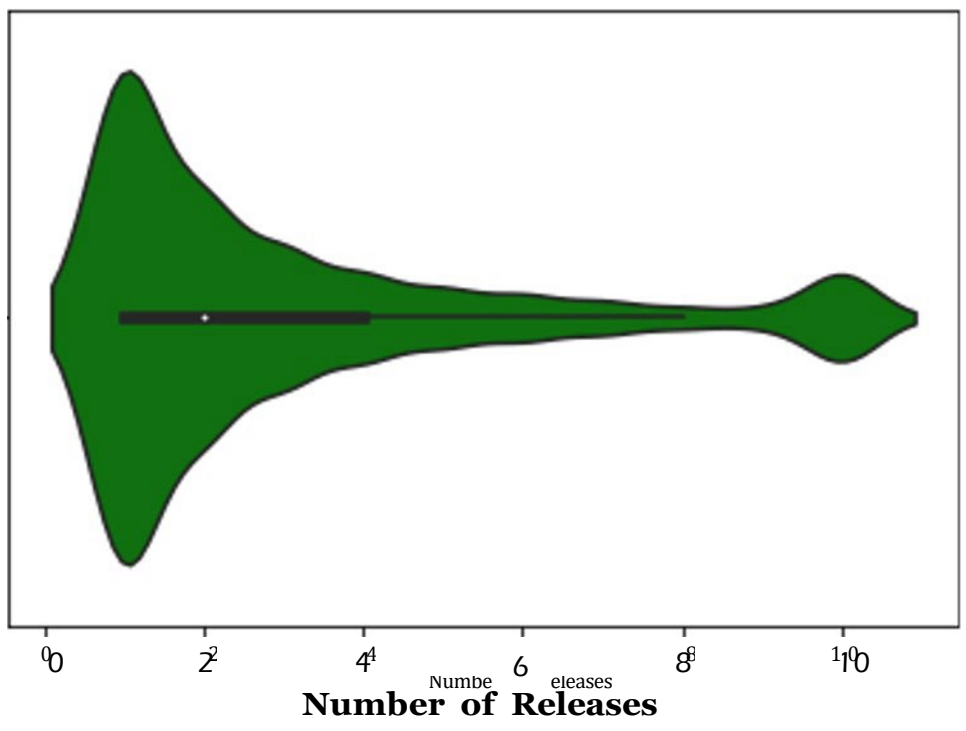

**Fig. 8** Number of releases vs number of actions in the marketplace

*Apps and actions each provide a set of unique attributes which can be quantified. In particular, majority of actions provide a detailed release history over multiple versions.*

## 4.3 Platform Attributes

Al-Subaihin et al. [1] have explored the impact of platforms on software engineering practices. While the study is only limited to the mobile app development, the impact of platforms on software is non-trivial. In this study, by 'platform attributes' we are referring to the attributes defined and provided by the GitHub Marketplace.

### *4.3.1 Developers*

The marketplace provides developer information for apps and actions. Overall, there are 5,928 Developers hosting their apps or actions in the marketplace. Among them, 368 published only apps, 5,509 published only actions, and 51 has both types of products on the platform. 20.70% (1,227) of the developers have more than one product (app or action) in the marketplace. *Azure* has the most number of products (One app and 35 Actions). This is also the highest number of actions by one provider in GitHub Marketplace. This is followed by *reviewdog* with 25 Actions but with no app on the marketplace. *Devbotsxyz* owns the highest number of apps with five apps but one action.

Figure 9 shows the number of unique developers working on apps and actions in each marketplace category. 'Continuous Integration' (C5), 'Utilities' (C3), and 'Deployment' (C10) are the top three categories that have the highest number of active developers. This does not come as a surprise as these three categories also have the most number of apps and actions in the marketplace (see Table 8). On average, each developer has 1.40 Products (apps or actions) on the marketplace. For each category, if we consider the ratio between number of developers to the number of products (i.e apps or actions), 'Continuous Integration' (C5, $ratio$ = 1.30), 'Utilities' (C3, $r$ = 1.29), and 'Code Quality' (C19, $r$ = 1.25) are ranked highest among all the categories. The very close ratios for all the categories implies that majority of the developers (79.30%) on the marketplace tend to publish only one tool regardless of the category.

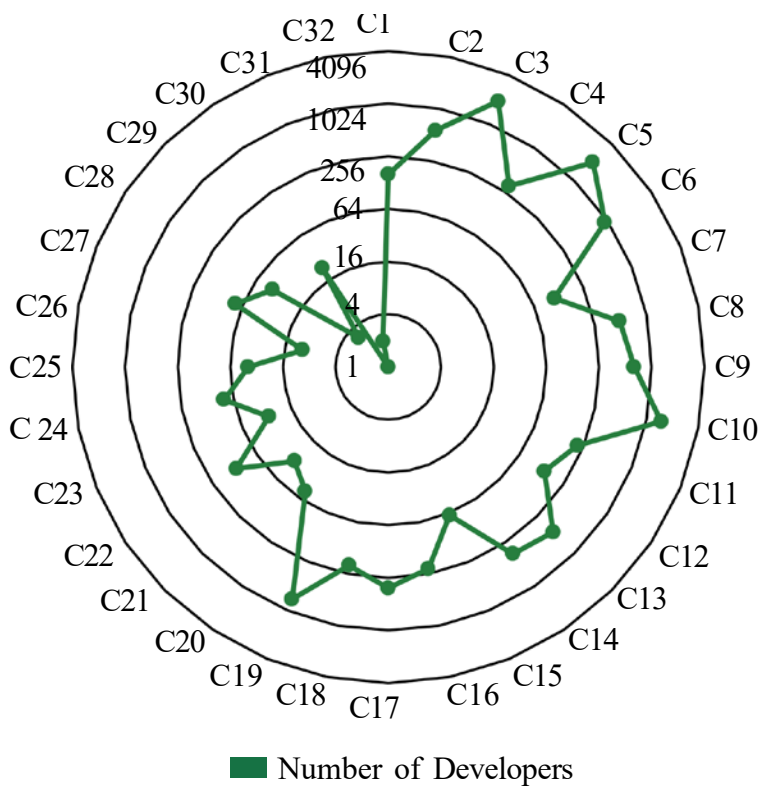

**Fig. 9** Number of developers (apps and actions) per category

#### *4.3.2 GitHub Verification of creators and products*

GitHub Marketplace has two separate processes for verifying creators and verifying products (i.e apps or actions).

To offer paid plans for apps or to include a marketplace badge in the app listing, the app developer must complete the publisher's verification process for their organization and become a 'verified creator'. As part of this process, the publisher must ensure that their basic profile information is accurately filled out, including an email address for support and updates from GitHub. Two-factor authentication also needs to be enabled for the organization. The organization must have verified ownership of their domain and ensure that a 'Verified' badge displays on the organization's profile page. GitHub will review the details and inform the organization once their publisher verification is complete. Once the organization has been verified, they can publish paid plans for their apps.

Actions with the *verified creator* badge indicate that GitHub has verified the creator of the action as a partner organization. However, there are two possible levels of verification for apps:

– App meets the requirements for listing: These apps meet the listing requirements but the publisher has not been verified. Apps with this badge cannot change their pricing plan until the publisher successfully applies for verification. Or,
– Publisher domain and email verified: These are granted to the apps that are owned by an organization that has verified ownership of their domain, confirmed their email address, and required two-factor authentication for their organization. In other words, the app went through the publisher verification process and was successfully granted a 'verified creator' status.

There are 3.67% (304 out of 8,225) Verified creators in the marketplace. 2.67% (209 out of 7,846) of the actions on the marketplace have verified creators. 69.09% (304 out of 440) of the apps in the marketplace are *not* verified by GitHub and does not have any level of verification badge. Among the 136 Verified apps, 41 has the 'App meets the requirements for listing' status while 95 holds 'Publisher domain and email verified' status.

#### *4.3.3 GitHub Recommended Apps*

15.0% (66 out of 440) of the apps on the GitHub Marketplace have a label stating *Recommended*. The number of installs for these apps are not provided. GitHub does not provide any official description of the eligibility criteria for this tag. We checked the list of recommended apps across different users as well as for three users over time. The set of recommended apps are static and is not curated based on a user or usage over time. Further, we hypothesized that this could be relevant to apps verifiability and authentications. However, our observations showed that these recommended apps have a lower percentage (18.18%) of verification. Only 22.72% of these recommended apps have a verified domain compared to 47.06% of the non-recommended apps with at least one verified domain. 'Localization' (33.33%), 'Time Tracking' (28.57%), and 'Support' (25.0%) are the categories with the highest proportion of recommended apps.

*Majority (79.3%) of the developers have published only one tool on the marketplace. Only 2.67% Action developers and 21.59% App developers are verified creators on the marketplace. Despite flexible pricing plans for the tools on the marketplace, only verified creators can distribute paid tools.*

## 4.4 User and Usage Attributes

In the GitHub Marketplace, the attributes provided for apps and actions are not similar. For instance, the actions have information about the 'number of stars' while for apps, the 'number of installs' are provided. We will refer to the apps with the fewest number of installs as the least popular apps. Similarly, the least popular actions are the ones with the fewest number of stars. The inaccuracy of these ratings have been discussed by Ruiz et al. [32] for mobile apps. In the context of our study, while we use the number of stars and the number of installs as the most accessible user feedback, it is essential that the reader considers the limitations to this interpretation.

### *4.4.1 Apps' Number of Installs*

Cracking the code of success for products is the ultimate wish of product owners. In the context of platform-mediated software products (particularly mobile app stores), success analysis has been discussed qualitatively by gathering developer [34, 1] and user perceptions [22] or mining and observing quantitative measures for extracting the confounding factors to the app's rating and number of installs [46]. Following that footstep, we performed a heuristic search to mine confounding factors for the popularity of products on the GitHub Marketplace. Due to GitHub's mechanism of masking the number of installs for the 'recommended apps' (see Section 4.3.3), we only based our analysis in this section on 85% of the apps (374 apps out of 440).

Figure 10-(a) is the Boxplot distribution of the 'number of installs' for apps. The highest number of installs was 1.5 Million which belonged to `GitHub Learning Lab` followed by Travis CI with 314,000 Installs. `Slack + GitHub` is the third most installed app with 207,000 Installs. Among the top quartile of most installed apps, three are developed by GitHub. The average number of installs per app on the marketplace is 7,530.12. However, 92.25% of apps have less than the average number of installs. The median number of installs is only 143.5. Among the apps, the highest average number of installs belong to categories 'GitHub Created' (with 594,500 Installs), 'Learning' (115,748.08), and 'API Management' (75,320).

### *4.4.2 Actions' Number of Stars*

GitHub users star repositories to keep track of the projects they find interesting as well as to show appreciation to the repository maintainer for their efforts. GitHub may recommend related content to a user on their personal dashboard based on their starred repositories. For this reason, users also star repositories in order to find similar projects on GitHub. Many of GitHub's repository rankings depend on the number of stars a repository has[17]. In our analysis, we also use the number of stars as a popularity metric for the actions on the marketplace.

The highest number of stars for actions belongs to `Super-Linter` developed by `github` with 6,800 stars followed by `yq - portable yaml processor` developed by `mikefarah` with 3,900 stars. Figure 10-(b) demonstrates the Boxplot distribution of the star ratings among actions. Azure has the most number of actions (35 Actions) followed by reviewdog (25 Actions). When looking into the proportion of number of actions in a category and the number of stars, 'Time tracking' and 'Community' are the top two categories while 'Game CI' and 'Backup Utilities' are the least popular. Also, 39 out of the 40 Actions with the least number of stars belong to the 'API Management' category.

[17] https://docs.github.com/en/get-started/exploring-projects-on-github/saving-repositories-with-stars

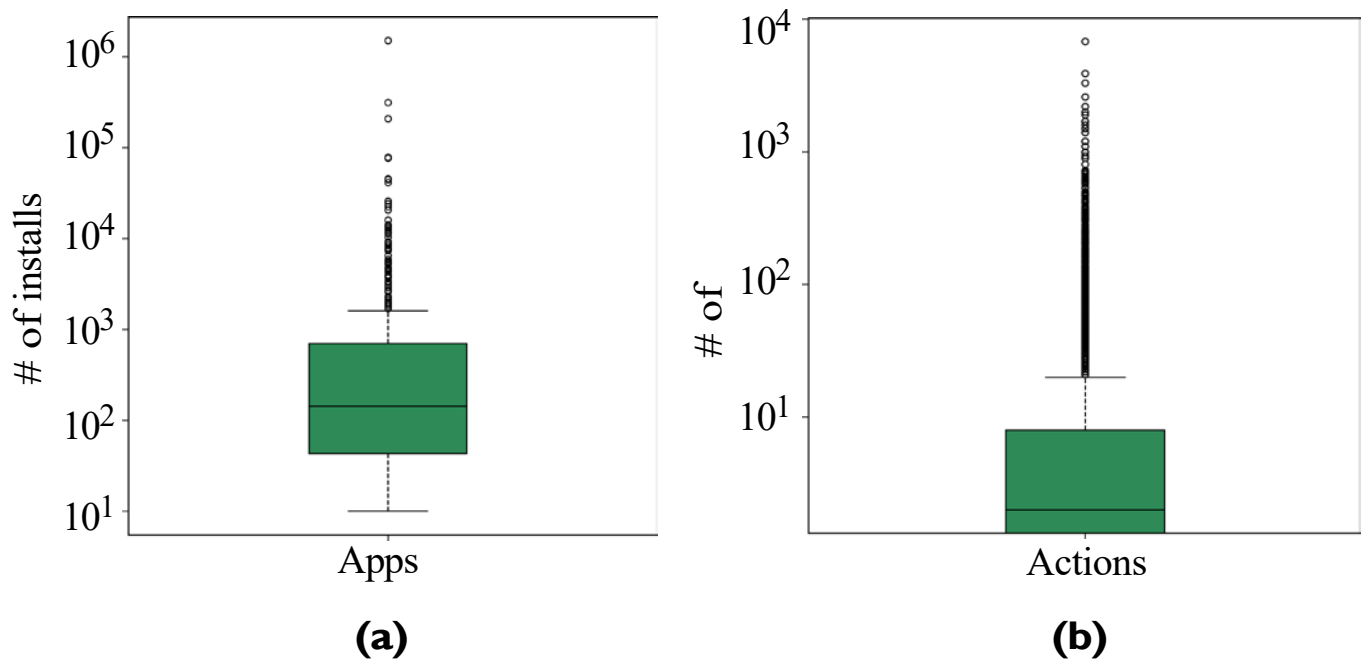

**Fig. 10** Distribution of number of installs for **(a)** apps and number of stars for **(b)** actions on the marketplace

### *4.4.3 Actions' Contributors*

Additionally, the marketplace also lists all the contributors of an action. Contributors are the software developers who open an issue, propose a Pull Request, or commit any type of change to the default branch of an action repository. A developer may or may not be a contributor. 7,111 Actions (90.63%) in the marketplace have at least one contributor. 61.81% actions have only one contributor. On the other hand, a total of 145 Actions have 12 Contributors, which is the largest number of contributors for actions on GitHub. 4,915 out of 7,878 Actions (62.4%) involved their developer as a contributor. However, we found a weak negative point-biserial correlation (r = -0.32) between the number of contributors and the action's developer also being a contributor.

*We measured the popularity of apps and actions based on the number of installs and the number of stars, respectively. The apps created by GitHub (C32) and the apps on 'Learning' (C27) are the top two app categories with the highest average number of installs. 'Time tracking' (C28) and 'Community' (C12) have the highest average number of stars among the actions.*

## 4.5 Relation Between the Attributes

66.8% (250) of the 374 Apps that display their number of installs are *not* verified by GitHub. The average number of installs for these apps is 436.7 compared to an average of 21,831.36 Installs for the apps that are verified by GitHub. Similarly, the average number of installs for the 191 Apps that have at least one verified domain is 11,711.18. On the other hand, the average installs for apps without a verified domain is only 2,336.41 for 248 Apps. This phenomenon also holds true for actions, although only 2.67% (209) of the actions on the marketplace have verified creators. The average number of stars for these 209 actions is 115, which is significantly more than 17.35, the average stars for actions without a verified creator. 2,199 Actions have 0 (zero) stars, of which only 0.14% (3 out of 2,199) have a verified creator.

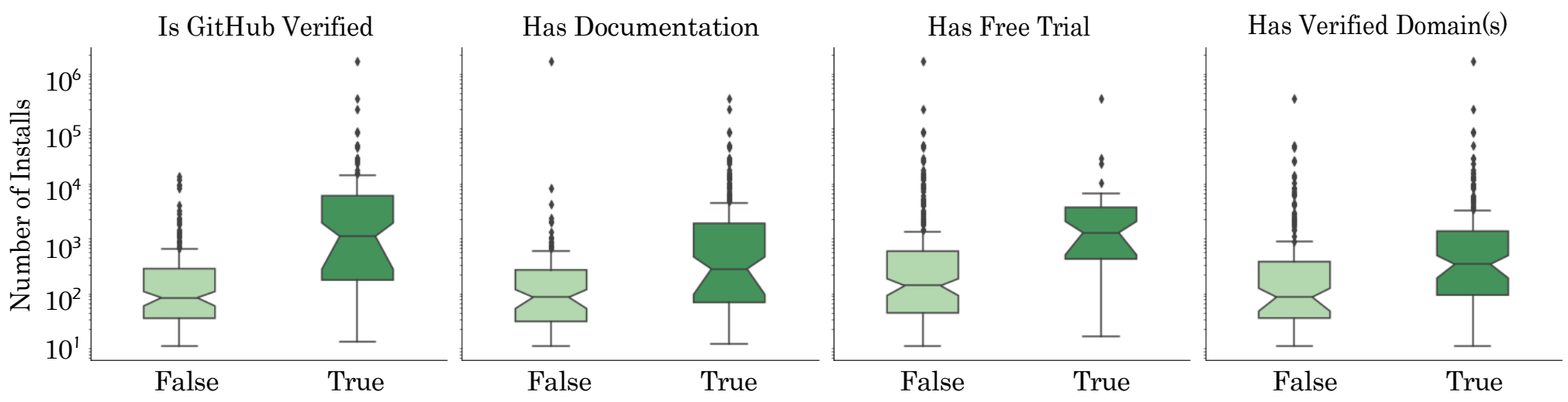

**Fig. 11** Distribution of the number of installs in regards to different marketplace properties for apps

One common thing about these least popular apps on the marketplace is their lack of verification. None of the ten least popular apps are verified by GitHub and only one of them has a verified domain, in contrast to the overall percentage of 43.51% for apps with verified domains. Three of these ten Apps belong to the 'Utilities' category. Only 20% among these ten apps have a link to the documentation, which is far below the overall percentage of 64.44% on the marketplace. Notably, none of the ten apps have a paid version. However, 79.68% of the apps on GitHub do not have a paid version. Therefore, this might not be a distinctive feature of the apps with the least number of installs. Figure 11 shows box-plot distributions of the number of installs in regards to different marketplace properties for apps. The median number of installs is higher for apps that are verified by GitHub compared to the apps that are not verified. Similarly, this median value is higher for apps that have documentation, free trial, or verified domains. However, we did not find any strong correlation between number of installs and these app properties. The point-biserial correlation coefficients between number of installs and (i) is GitHub verified (0.12) (ii) has documentation (-0.04) (iii) has free trial (0.01) (iv) has verified domain(s) (0.06) are trivial.

Verification is a contributing factor for higher number of stars for the actions as well. Only 2.66% Actions have a verified creator on the marketplace. However, 24 out of the top 100 Actions with the most number of stars have a verified creator. Five out of the top ten actions belong to the 'Utilities' category. Three among these ten actions share the categories 'Deployment' and 'Publishing'.

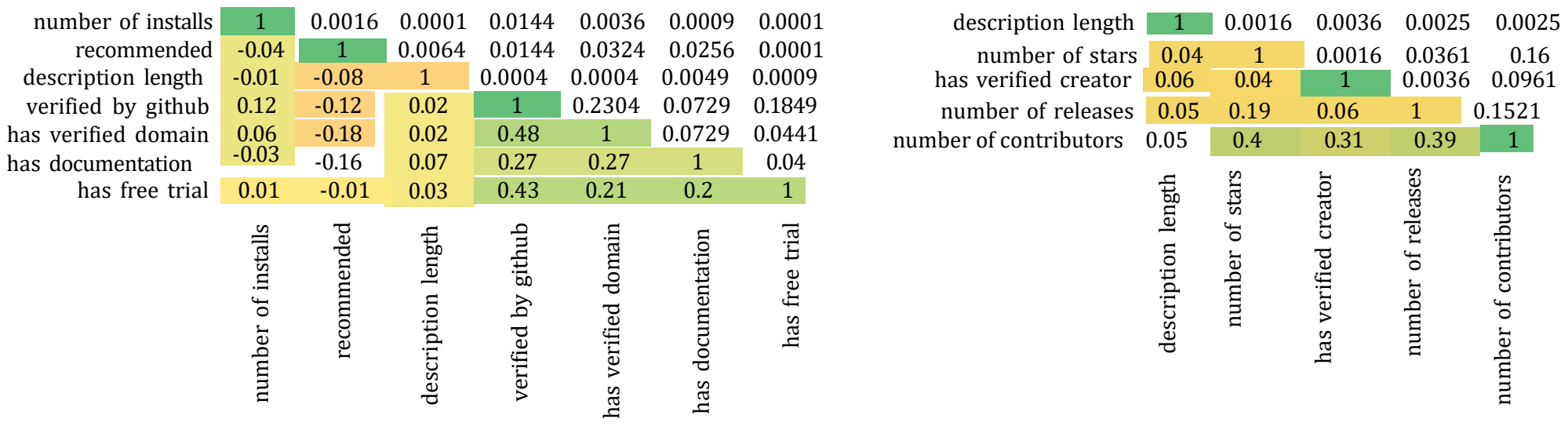

**(a) Apps** **(b) Actions**

**Fig. 12** Correlation matrix between different attributes for (a) apps and (b) actions. The upper triangle shows the effect size. As effect size, we use the coefficient of determination (goodness of fit).

*We analyzed the relationships between a number of attributes and the popularity of apps and actions. Among all, the verification status demonstrated the most strong correlation.*

## 5 RQ2: Mapping Studies of Studies in Open Source Software Engineering

In the previous section, we explored the status quo of the GitHub Marketplace, which are the tools practitioners use in the open source ecosystem (**RQ1**). To address the knowledge mobilization gap (**RQ3**), we performed a systematic mapping of software engineering studies in **RQ2**. We discuss the results of the mapping study in this section.

As discussed in Section 3.2, we gathered relevant papers on automation within the open source software ecosystem between the years 2000 to 2021. More importantly, we were interested in gathering and synthesizing the status of automation techniques and tools offered for the open source community. Recent years have seen an increase in the number of articles published in this field. Figure 13 shows the number of publications per year relevant to 'automation' and 'open source software engineering' in the top 20 software engineering venues. There is a steep increase in the number of publications starting from the year 2009, and there is another peak in 2015 onward. There were 32 publications in 2021, compared to only six publications in 2011 (ten years period), strengthening the case that researchers are recently more active in this domain than ever before.

Following the systemic protocol procedure described earlier (see Section 3.2), two authors independently mapped the papers included in the study into the app and action categories (32 Categories overall). Table 6 demonstrates the results of this mapping.

### 5.1 Quantitative Analysis of the Mapped Studies

As shown in Figure 14, 'Code quality' was the most popular category in software engineering literature with 97 out of 292 papers. We also examined how often two categories appeared together and how many papers the categories involved. Figure 16 demonstrates the frequency of categories chosen together for papers. The biggest intersection belongs to 'Code quality' and 'Code search' with 24 Papers sharing these two categories. This is followed by 'Code quality' and 'Code review' with 20 Papers. 'Utilities' pairs with 22 out of 26 other categories for papers, followed by 'Code quality' (with 20 other categories), and 'Support' (with 20 other categories).

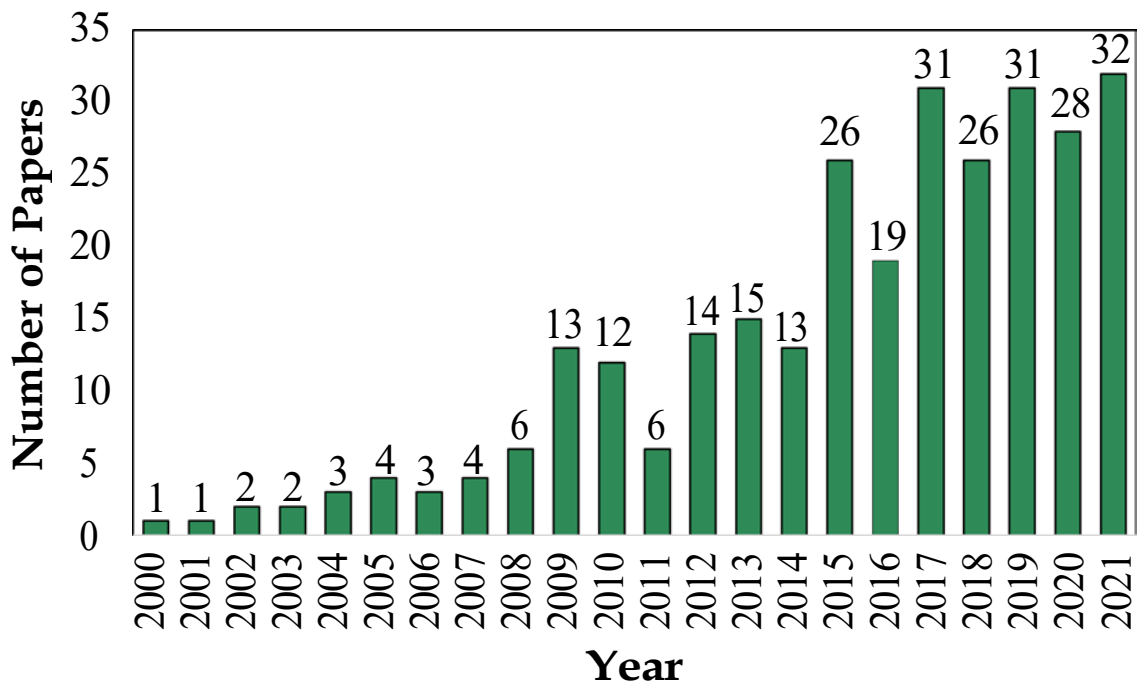

**Fig. 13** Number of categorized papers within our systematic mapping study (**RQ2**) in each year

**Table 6** Papers mapped to GitHub Marketplace categories

| Category | ID |
|---|---|
| API management and Checking | P16, P29, P64, P140, P186, P261, P340, P361 |
| Testing | P14, P18, P28, P60, P71, P76, P79, P83, P89, P91, P92, P95, P107, P114, P128, P130, P131, P132, P137, P138, P160, P164, P168, P170, P173, P176, P177, P179, P185, P188, P192, P193, P203, P209, P217, P221, P222, P227, P249, P260, P262, P263, P264, P266, P273, P276, P277, P279, P282, P283, P284, P288, P303, P323, P331, P341, P350, P357, P360, P363, P364 |
| Utilities | P8, P13, P36, P38, P42, P51, P62, P64, P73, P100, P103, P108, P117, P120, P130, P139, P140, P150, P151, P154, P155, P159, P163, P164, P166, P168, P172, P175, P189, P194, P195, P207, P231, P236, P259, P267, P270, P301, P340 |
| Reporting | P21, P44, P45, P50, P65, P67, P69, P71, P81, P99, P100, P105, P109, P174, P180, P182, P216, P220, P234, P239, P240, P287, P327, P334, P350, P359, P364 |
| Continuous integration | P37, P83, P98, P194, P224 |
| Publishing | P6, P25, P66, P98, P121, P186, P194, P215, P238, P242, P289, P291, P292, P304, P305, P313, P355, P356 |
| Support | P8, P15, P24, P28, P42, P48, P55, P72, P77, P80, P82, P84, P87, P98, P101, P105, P107, P117, P119, P121, P126, P135, P157, P168, P169, P170, P175, P197, P207, P215, P216, P229, P240, P242, P248, P265, P267, P272, P275, P286, P290, P294, P301, P305, P311, P328, P345 |
| Project management | P5, P6, P10, P15, P25, P37, P47, P57, P69, P88, P116, P121, P127, P135, P148, P152, P158, P182, P187, P238, P244, P248, P252, P256, P257, P259, P265, P269, P271, P278, P286, P289, P290, P291, P292, P298, P300, P313, P322, P330, P334, P344, P345, P352, P355, P356, P359 |
| Code review | P1, P2, P22, P29, P36, P51, P59, P68, P70, P78, P80, P82, P93, P99, P104, P106, P111, P117, P122, P125, P133, P140, P144, P155, P156, P165, P166, P187, P205, P230, P233, P247, P250, P289, P299, P314, P317 |
| Deployment | P10, P79, P163, P189, P291, P292, P311, P326 |
| Chat | P35, P39, P74, P220, P236 |
| Community | P5, P25, P27, P35, P37, P39, P47, P50, P59, P72, P73, P74, P94, P113, P118, P119, P134, P242, P275, P305 |
| Dependency management | P7, P17, P78, P143, P159, P161, P171, P180, P193, P196, P201, P202, P215, P221, P225, P228, P230, P235, P243, P253, P254, P255 |
| AI Assisted | P10, P17, P22, P36, P56, P61, P65, P75, P81, P92, P102, P112, P127, P136, P142, P146, P147, P153, P155, P157, P158, P161, P163, P166, P171, P182, P195, P212, P225, P227, P228, P232, P235, P244, P248, P249, P250, P252, P253, P254, P255, P256, P262, P266, P290, P298, P300, P307, P311, P316, P325, P336, P347, P354 |
| Open Source management | P14, P29, P47, P57, P72, P73, P74, P77, P83, P84, P94, P108, P118, P119, P148, P220, P223, P236, P244 |
| Security | P55, P136, P148, P172, P191, P196, P201, P279, P299, P315, P361 |
| Monitorin | P80, P110, P116, P124, P132, P139, P143, P151, P152, P158, P167, P171, P174, P177, P196, P252, P265, P270, P278, P326 |
| Code quality | P1, P2, P16, P28, P44, P60, P62, P68, P70, P75, P78, P89, P90, P93, P95, P97, P101, P102, P103, P104, P106, P109, P111, P112, P114, P120, P122, P124, P125, P127, P129, P133, P137, P139, P142, P143, P144, P145, P146, P147, P150, P151, P153, P154, P156, P157, P160, P165, P167, P170, P172, P178, P180, P181, P184, P187, P188, P189, P193, P195, P203, P204, P206, P214, P217, P223, P224, P229, P230, P231, P238, P247, P254, P258, P266, P268, P272, P273, P278, P279, P281, P287, P288, P293, P302, P306, P309, P312, P313, P314, P317, P323, P330, P332, P346, P351, P365 |
| Localization | P16, P145, P146, P147, P150, P161, P202, P221, P225, P228, P235, P253, P255 |
| Desktop tools | P4, P100, P108, P169, P240 |
| Mobile | P50, P76, P79, P109, P128, P183, P260 |
| IDEs | P42, P51, P205 |
| Mobile CI | P76 |
| Code search | P1, P7, P14, P22, P27, P44, P64, P68, P70, P90, P93, P95, P101, P104, P106, P112, P114, P120, P125, P133, P137, P153, P154, P156, P164, P165, P185, P197, P209, P212, P239, P264, P268, P272, P273, P302 |
| Code scanning ready | P2, P90, P102, P142, P145, P202, P209, P216, P239, P267, P288, P299 |
| Learning | P5, P45, P122, P126, P174, P205, P275, P294, P327, P364 |
| Time tracking | P35, P57, P152, P250 |

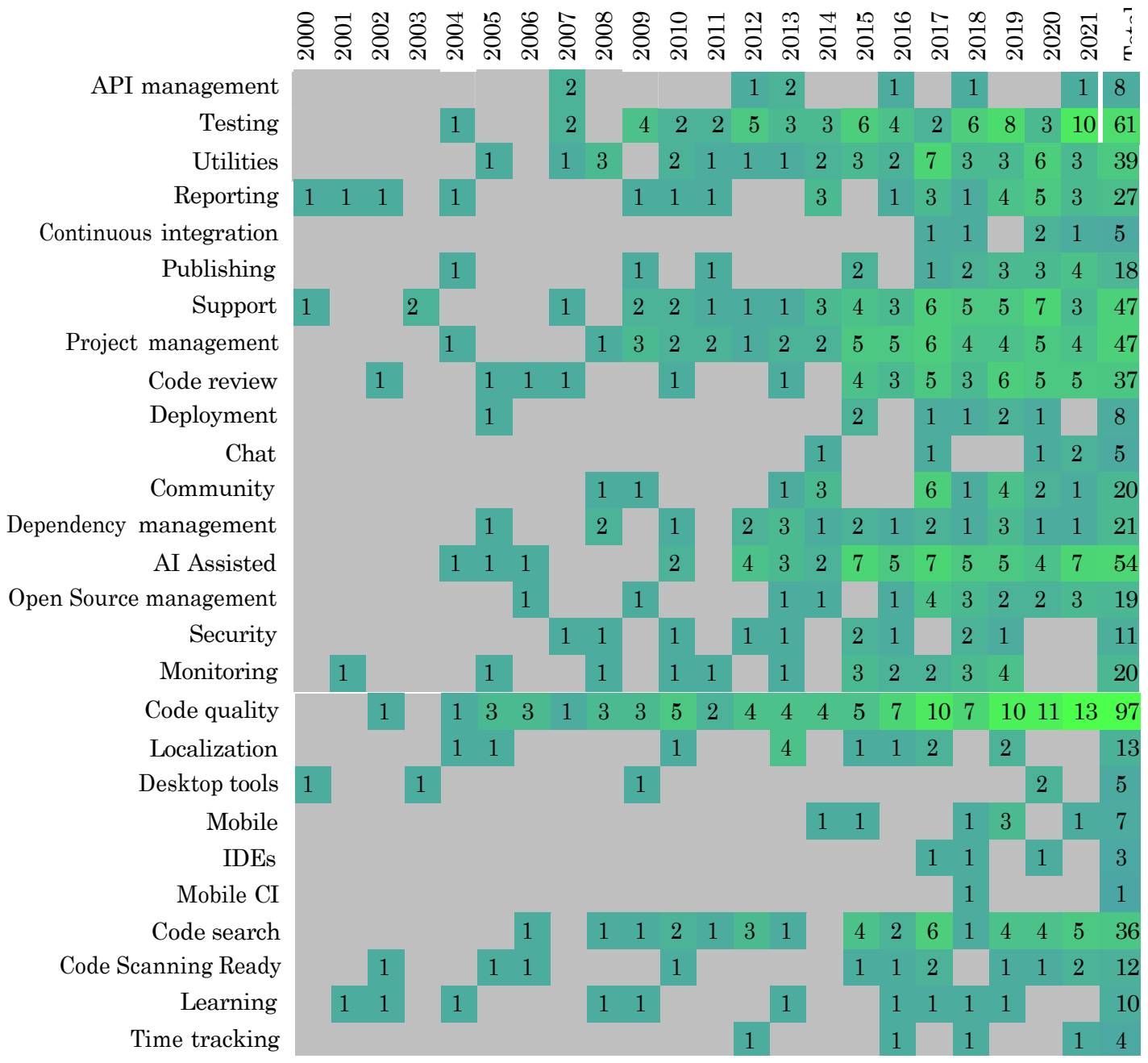

**Fig. 14** Number of papers in each year per category. Categories with zero number of paper were excluded from the heatmap.

On the other hand, 'Mobile CI' pairs with only two other categories while 'IDEs' intersect with four other categories and 'Time tracking' appear together with five other categories.

We looked into the number of distinct authors per category as an indicator for the number of contributors in each domain. We counted a total of 871 distinct authors for 292 Papers in our mapping study. Figure 15 shows the number of individual authors in each marketplace category. 'Code quality' (C19), 'Testing' (C2), and 'AI Assisted' (C15) are the top three categories that have the highest number of active researchers. These are also the top three categories with the most number of papers (see Table 8)[18]. Similarly, the category 'Mobile CI' (C24) has the least number of papers (one) as well as distinct authors (three). 25.68% of the papers (75 out of 292) have three authors while the median number of authors per paper is 3.5. 13 Papers out of 292 (4.45%) have only one author.

*'Code quality' has the highest number of mapped studies among all categories. Papers in this category also appear frequently in 'Code search' (24 times) and 'Code review' (20 times). 'Code quality' also has the highest number of (299) distinct authors, followed by 'Testing' (208) and 'AI Assisted' (206).*

[18]There is a strong positive correlation (r = 0.99) between the number of papers and the number of unique authors researching in a category across all the papers.

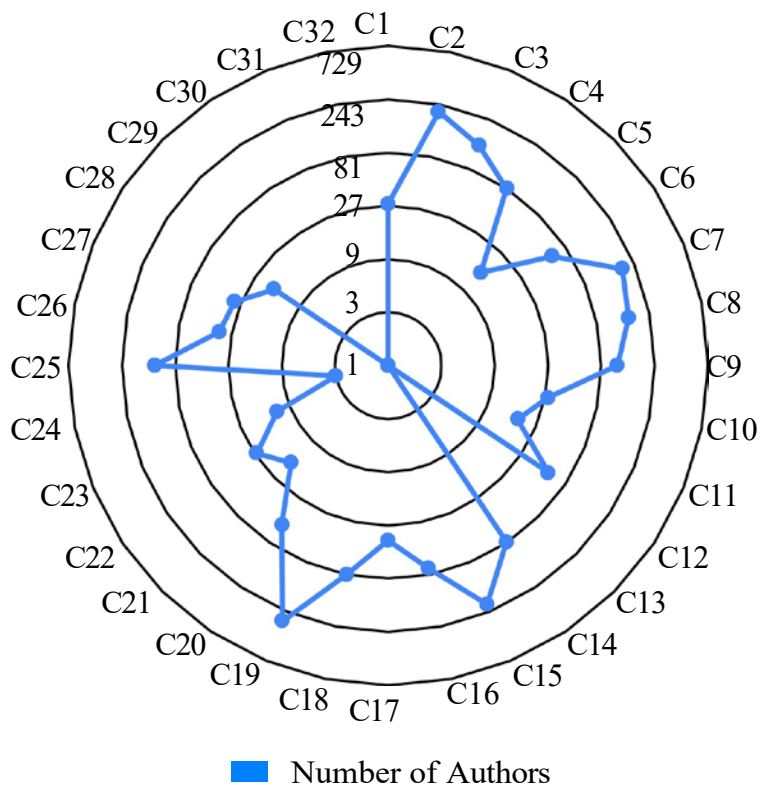

**Fig. 15** Number of authors per category

## 5.2 Qualitative Analysis of the Mapped Studies

Software programs often do not perform in isolation. GitHub Marketplace defines 'API management' (C1) as the tools to structure API infrastructure to enable various internet gateways to interact with offered services. On the other side, 'Dependency Management' (C14) tools help secure and manage third-party dependencies. Literature in both categories discuss approaches and tools for bug detection and requirement traceability.

The marketplace encourages developers to categorize tools intended to help them eliminate bugs and ship their products in the 'Testing' (C2) category. The category involves test management tools, logging tools, and benchmarking tools. Further, the 'Code Scanning Ready' (C26) category is dedicated to static analysis, dynamic analysis, container scanning, linting, and fuzzing tools. These tools are mostly integrated with GitHub Code Scanning SARIF[19] Upload. Our mapping resulted in 61 Studies that suggested automation in the 'Testing' category, and 18 of them offered tool support. Others focused on proposing models and algorithms or benchmarking the performance of different techniques for improved testing practices and tested these models using open source projects.

GitHub Marketplace offers 'Utilities' (C3) as a set of auxiliary tools to enhance the developers' experience on GitHub. The category includes tools for backup, bots, pull request automation, backup tools, visualization tools, and environment setup tools. We mapped 39 Studies in the 'Utilities' category and similar to the marketplace products, these studies are distributed across different phases of the software life cycle. The 'Monitoring' (C18) category provides tools to monitor the impact of code changes and help developers to measure performance, track errors, and analyze applications. We mapped 20 Studies in this category which discuss topics such as crash localization, crash reproduction, reporting problems, code-smell detection, among others. When it comes to 'Reporting' (C4), the GitHub Marketplace consists of tools that provide insight into the team's development activities. We categorized 41 Papers into the 'Reporting' category. This category involves variety of studies which assist the developers to get further insight into their project through different medias such as measurement or visualization. However, majority of the studies that fell into this category were mainly focused on mining repositories and emphasizing the need for automation tools.

[19]SARIF stands for Static Analysis Results Interchange Format

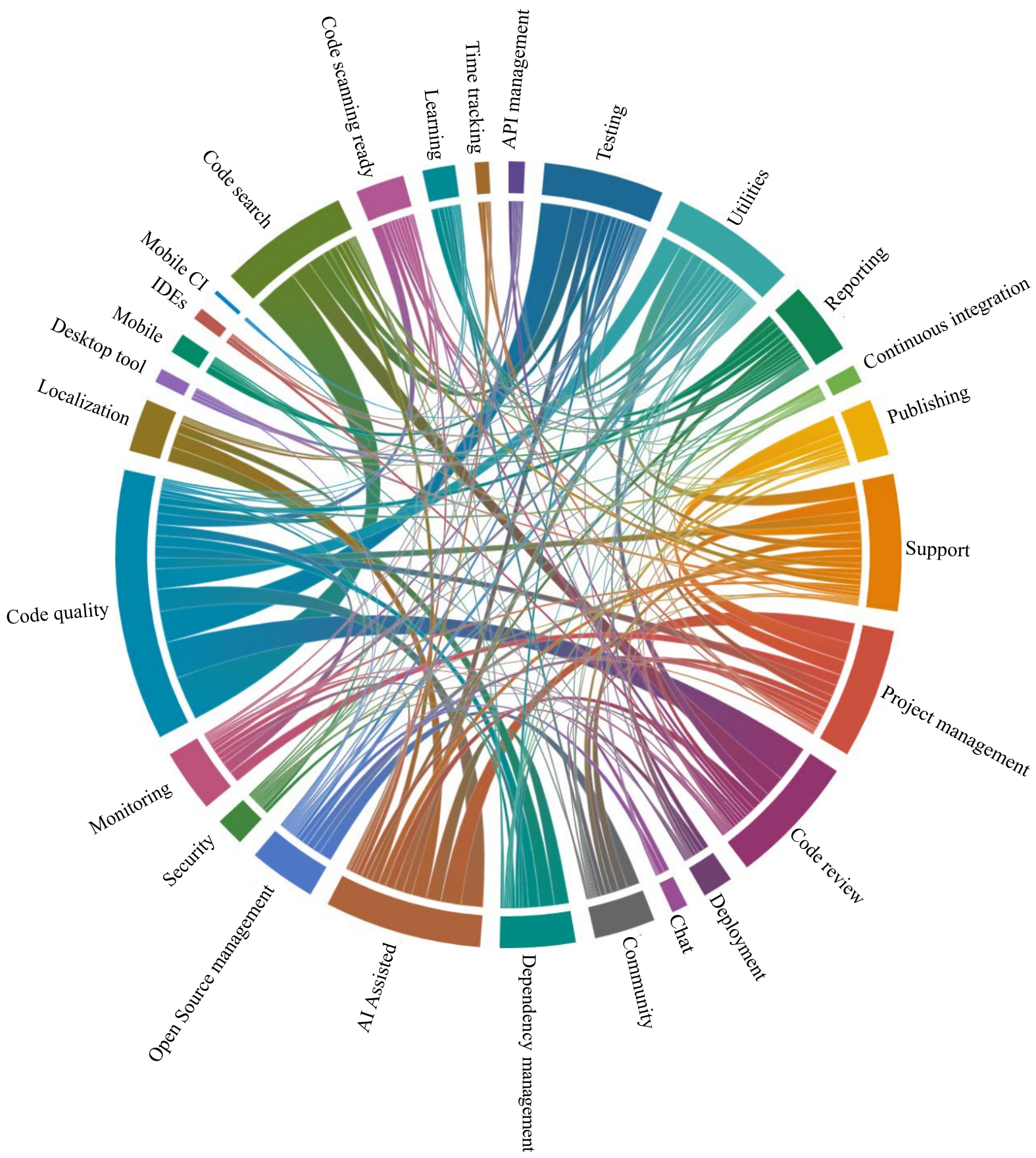

**Fig. 16** Frequency of categories appearing together for the papers in our analysis.

To automatically build and test code when pushing it to GitHub and preventing bugs from being deployed to production, the GitHub Marketplace introduces 'Continuous Integration' (C5), with subcategories 'Container CI' (C13), 'Mobile CI' (C24), and 'Game CI' (C29). Within our mapping of papers relevant to 'Continuous integration', we only considered studies directly relevant to containers, mobile, or games as relevant to each subcategory C13, C24, or C29. Our search did not result in any publications within the defined protocol for continuous integration of containers or open source games.

In 2015, Vasilescu et al. [47] tapped into the opportunity of accessing various projects on GitHub, which were at different stages of process integration and automation, to pick out the effect of continuous integration. Their study showed that by using continuous integration, the core developers of each team were more effective at merging pull requests and finding more bugs. Since then, a number of studies have focused on this topic. However, when it comes to automation of the continuous integration process for open source software, we retrieved and mapped five papers since 2017 (see Figure 14 for the yearly numbers).

Marketplace's 'Publishing' category (C6) consists of tools for getting sites ready including release tools for production while 'Deployment' (C10) tools refer to streamline code deployments. We mapped 18 Papers in the 'Publishing' category and only eight papers being primarily relevant to the 'Deployment' category. Studies from both categories discuss cloud deployment and evaluating deployments. Similar to the marketplace, the mapping in state of the art has a high overlap with the 'Continuous Integration' (C5) and 'Testing' (C2) categories. Among them, `CCBot` [6] is a tool that uses static analysis to detect faults at runtime and guarantees the developer can deploy (or test) the code without additional manual effort. The rest either offered a framework or algorithm for deployment and the analysis stays non pragmatic.

When it comes to community support, GitHub provides a set of tools. The apps and actions in the GitHub's 'Support' (C7) category are designated to help with the needs of the team and the customers. 'Localization' (C20) consists of tools to extend software's reach[20] and to localize and translate continuously from GitHub. The tools in the 'Chat' (C11) category, in particular, are designed to bring GitHub into the conversations. This is while the marketplace's definition of 'Community' (C12) tools is quite brief as 'Tools for the community'. We also decided to briefly describe studies relevant to the 'Learning' (C27) category along with these as they support individuals to get the skills they need for leveling up. In our mapping study, we identified two main categories among the studies where the focus is either on adopting the engineering process to support the end users or to support the productivity of the development team.

To assist software teams in organizing, managing, and tracking projects, GitHub Marketplace offers tools in the 'Project management' (C8) category that build on top of issues and pull requests. Also, GitHub acknowledges the difficulties teams might face when managing their open source projects and provides tools under 'Open Source management' (C16) to make managing more 'fun and manageable'[21]. Our analysis resulted in mapping 47 Studies in the 'Project management' (C8) category. Similar to the marketplace, the items within this category are of a diverse nature. Among them, a few are particularly addressing automation tools for developers in the open source environment. Only five studies are particularly proposing a tool set while the rest are offering insight on the importance of the automation in the context of project management. The marketplace offers 'Time Tracking' (C28) tools for tracking progress and predicting the length of tasks based on team's coding activities. We mapped four studies in this category which mainly discuss reducing code review effort and documentation effort, prioritizing documentation, and issue management i.e. detecting duplicate issues.

GitHub has a specified category, 'AI Assisted' (C15), for the tools that are enabled by artificial intelligence (AI). While many studies in the software engineering literature are using components of machine learning, we discuss the ones that we primarily mapped to this category. These studies have focused on bug localization, automatic bug reporting, finding traceability links, vulnerability detection, automatic code generation, to name a few areas. To find, fix, and prevent security vulnerabilities, the marketplace offers 'Security' (C17) tools. Following the same direction for literature, we mapped a number of studies that offered tools for vulnerability detection, code analysis, finding and fixing broken dependencies, and more.

GitHub Marketplace defines 'Code Quality' (C19) as the category to 'Automate your code review with style, quality, security, and test-coverage checks when you need them.' 'Code Review' (C9) is defined as the set of tools that ensure code meets quality standards, resulting in developers shipping software with confidence. Along the same line, 'Code search' (C25) is the category dedicated to the tools that query, index, or hash the semantics of source code. We mapped 97, 37, and 36 Papers in C19, C9, and C25, respectively. Literature in these areas focus on static and dynamic analysis, automatic refactoring, test coverage, bug localization and reporting, program comprehension, traceability link detection, technical debt analysis, among other tasks. 'Desktop Tools' (C21) are offered by the marketplace to assist developers to run tools natively on their local machines. The 'Mobile' (C22) tools are designed to improve the workflow for the small screen. 'IDEs' (C23) is a category of tools that help developers in finding the right interface to build, debug, and deploy their source code. These are relatively less popular categories in literature. We mapped five, seven, and three studies in C21, C22, and C23 respectively.

[20]This should not be mistaken with the bug localization in software research.
[21]https://github.com/marketplace/category/open-source-management

We extracted bigrams from paper titles and abstracts for all the papers in a category. These bigrams are presented in Table 7. We followed the same protocol and process as for extracting bigrams for tools on the marketplace (see Table 5). These bigrams represent the key features or phrases.

One of the common features in literature is bug reporting, appearing in seven categories namely 'AI Assisted' (32 Mentions of this feature), 'Project management' (32), 'Code quality' (24), 'Reporting' (22), 'Code search' (14), 'Code review' (11), and 'Support' (8). Another feature that is widely shared by multiple categories is requirement traceability. 'Code review' (10), 'Dependency management' (10), 'Utilities' (10), 'IDEs' (10), 'AI Assisted' (9), and 'Localization' (8) all have papers on the topic of requirement traceability. The categories 'Code quality' (7), 'Code Scanning ready' (6), 'Utilities' (6), and 'Deployment' (4) are similar in that they all have papers that facilitate static analysis. The topic of test automation is discussed in papers that fall under the categories 'Testing' (13), 'Mobile' (4), 'Mobile CI' (4), and 'Open source management' (2). Table 7 shows the top features per category along with the frequency in which these features appear in paper titles and abstracts.

*Studies in different categories often discuss similar topics/features. The most common feature in literature is 'bug reporting' (appearing in seven categories) followed by 'requirement traceability' (six categories), 'static analysis' (four categories), and 'test automation' (four categories).*

## 6 RQ3: The Gap

We performed a systematic study of the marketplace and the literature to perform a gap analysis in **RQ3**. The category 'GitHub Created' is excluded from this analysis as this category only applies to actions and not apps or papers.

### 6.1 The Interest of Researchers and Practitioners Across Categories

There are 871 distinct authors and 5,937 distinct developers in our dataset of 292 Papers and 8,318 GitHub products (440 Apps and 7,878 Actions). We presented the number of developers (Figure 9) and the number of authors (Figure 15) per category in **RQ1** and **RQ2**, respectively. By comparing the two, we found that researchers and practitioners are interested in different subject categories, as shown in Figure 17. 23.65% (206) Researchers are publishing in 'AI Assisted', which is the third most popular category in terms of number of authors per category. In contrast, only 1.15% (68) Developers are building apps and actions in this category. The Percentage Point (pp) difference is 22.5. We have similarly striking gaps between the percentage of authors and developers working in the 'Code quality' (21.8 pp more authors) and 'Support' (19.43 pp more authors) categories. 20 Categories out of 32 have a higher percentage (ranging from 0.54 pp to 22.51 pp higher) of authors compared to the percentage of developers working in those categories. On the other hand, only 1.72% (15) of the authors are working in the 'Continuous integration' category compared to 34.23% (2,032) of the developers contributing to the marketplace in this category. Glaring gaps such as this, where a higher percentage of developers are involved compared to researchers, also exist in other categories such as 'Deployment' (22.29 pp more developers) and 'Utilities' (17.03 pp more developers). Categories with the least difference between the proportion of researchers and developers are 'API Management' (0.54 pp more researchers), 'IDEs' (0.87 pp more researchers), 'Mobile' (0.91 pp more researchers), and 'Mobile CI' (1.04 pp more developers). Figure 18 shows the percentage point difference between the contribution of researchers and practitioners for each category. For eight categories, there is more than 10 Percentage Point (PP) difference between the percentage of researchers and practitioners.

**Table 7** Top features in each category (extracted from paper titles and abstracts). The numbers in parenthesis stand for the frequencies of each feature appearing in literature.

| Category | Top features |
|---|---|
| AI Assisted | bug report (32), fault localization (13), traceability link (9), vulnerability detection (6), code generation (5) |
| API management | code search (5), natural-language query (5), api documentation (3), api specification (2), api library (2) |
| Chat | developer discussion (4), issue report (4), bug fix (4), communication challenge (2), gain insight (2) |
| Code quality | technical debt (26), bug report (24), fault localization (12), automate refactoring (11), static analysis (7) |
| Code review | technical debt (16), program comprehension (14), bug report (11), trace link (10), code analysis (8) |
| Code scanning ready | static analysis (6), dynamic analysis (6), defect identification (5), statistical debug (4), security vulnerability (4) |
| Code search | technical debt (24), bug report (14), bug localization (10), mutation analysis (8), natural-language query (5) |
| Community | open source (9), manage assignee (7), follow recommendation (6), developer discussion (4), issue report (4) |
| Continuous integration | CI build (9), CI specification (5), automate build (2), failure prediction (2), model develop (2)) |
| Dependency management | bug detection (9), traceability link (8), verification validation (7), co-change dependency (4), requirement traceability (2) |
| Deployment | cloud deployment (5), static analysis (4), deployment evaluation (4), code generation (3), mobile apps (3) |
| Desktop tools | uml profile (5), case tool (3), modelling language (2), support editor (2), embed system (2) |
| IDEs | program comprehension (13), trace link (10), link creation (5), data platform (2), ide plug-in (2) |
| Learning | program comprehension (12), open source (3), similarity measurement (2), developer study (2), knowledge acquisition (2) |
| Localization | bug detection (9), verification validation (7), traceability link (6), bug track (2), requirement traceability (2) |
| Mobile | mobile apps (13), android apps (6), automate test (4), gui test (4), visual test (3) |
| Mobile CI | mobile apps (6), automate test (4) |
| Monitoring | severity label (6), problem report (6), code-smell detection (5), crash reproduction (5), crash localization (3) |
| Open Source management | issue report (9), mutation analysis (8), identify duplicate (5), bug fix (4), automate test (2) |
| Project management | bug report (32), bug assignment (13), resolve merge-conflict (6), improve maintainability (5), support tool (4) |
| Publishing | CI build (9), CI specification (5), cloud deployment (5), release schedule (4), deployment evaluation (4) |
| Reporting | bug report (22), log change (9), tag recommendation (5), program repair (4), tool support (3) |
| Security | vulnerability detection (6), code analysis (5), static verification (3), refactoring edits (3), broken dependency (2) |
| Support | CI build (9), bug report (8), analysis tool (5), bug localization (5), code analysis (4) |
| Testing | test suite (34), fault localization (33), mutation analysis (13), automate test (13), unit test (10) |
| Time tracking | duplicate issue (8), documentation effort (5), issue report (5), prioritize documentation (3), review effort (3) |
| Utilities | trace link (10), static analysis (6), tool support (5), code search (5), code design (3) |

Besides having a developer, most actions (90.63%) in the marketplace have one or more contributors. 64.49% of these contributors are working in the 'Continuous integration' category compared to 1.72% of the authors. Categories 'Utilities' (39.7 pp), 'Deployment' (38.71 pp), and 'Publishing' (17.82 pp) have considerably more percentage of contributors compared to authors. On the other hand, 'AI Assisted' (22.36 pp), 'Support' (18.87 pp), and 'Code search' (13.34 pp) have a higher percentage of authors than contributors on the marketplace.

Although there are 30 Action categories in the marketplace (at the time of writing this paper), GitHub actions was first introduced as an alternative to CI/CD services for GitHub repositories [10]. For this reason, it comes as no surprise that 'Continuous Integration', and 'Deployment' are two of the top three categories based on the number of developers. The other category, 'Utilities', also includes marketplace products related to CI/CD. These three categories also pair frequently between them for the products in the marketplace (see section 6.4). This results in a lot of developers being involved in these three categories. While the practitioners are focusing on getting their products on the marketplace as quickly and seamlessly as possible, thanks to the abundance of CI/CD products, the researchers are concerned more about facilitating development with the help of AI assisted technologies, improved code quality, and support to both the development team and the end users.

**Table 8** Ranking of categories based on the number of apps, actions, and papers in that category (sorted by number of papers)

| ID | Type | Category | # Apps | # Actions | # Papers | $Rank_{Apps}$ | $Rank_{Actions}$ | $Rank_{Papers}$ |
|---|---|---|---|---|---|---|---|---|
| C19 | Both | Code quality | 65 | 864 | 97 | 5 | 5 | 1 |
| C2 | Both | Testing | 26 | 662 | 61 | 10 | 7 | 2 |
| C15 | Both | AI assisted | 26 | 58 | 54 | 11 | 22 | 3 |
| C8 | Both | Project management | 80 | 507 | 47 | 3 | 9 | 4 |
| C7 | Both | Support | 12 | 126 | 47 | 19 | 19 | 5 |
| C3 | Both | Utilities | 118 | 2,388 | 39 | 1 | 2 | 6 |
| C9 | Both | Code review | 81 | 684 | 37 | 2 | 6 | 7 |
| C25 | Both | Code search | 5 | 37 | 36 | 25 | 24 | 8 |
| C4 | Both | Reporting | 27 | 315 | 27 | 9 | 12 | 9 |
| C14 | Both | Dependency management | 20 | 401 | 21 | 13 | 10 | 10 |
| C18 | Both | Monitoring | 28 | 188 | 20 | 8 | 15 | 11 |
| C12 | Action | Community | N/A | 155 | 20 | N/A | 17 | 12 |
| C16 | Both | Open source management | 25 | 244 | 19 | 12 | 13 | 13 |
| C6 | Both | Publishing | 18 | 1,058 | 18 | 14 | 4 | 14 |
| C20 | Both | Localization | 9 | 46 | 13 | 22 | 23 | 15 |
| C26 | Both | Code scanning ready | 3 | 7 | 12 | 28 | 29 | 16 |
| C17 | Both | Security | 43 | 345 | 11 | 6 | 11 | 17 |
| C27 | Both | Learning | 16 | 66 | 10 | 15 | 21 | 18 |
| C10 | Both | Deployment | 35 | 1800 | 8 | 7 | 3 | 19 |
| C1 | Both | API management | 15 | 176 | 8 | 16 | 16 | 20 |
| C22 | Both | Mobile | 10 | 135 | 7 | 20 | 18 | 21 |
| C5 | Both | Continuous integration | 76 | 2,554 | 5 | 4 | 1 | 22 |
| C21 | Both | Desktop tools | 14 | 20 | 5 | 17 | 28 | 23 |
| C11 | Both | Chat | 12 | 224 | 5 | 18 | 14 | 24 |
| C28 | Both | Time tracking | 7 | 33 | 4 | 23 | 25 | 25 |
| C23 | Both | IDEs | 10 | 24 | 3 | 21 | 26 | 26 |
| C24 | Both | Mobile CI | 6 | 91 | 1 | 24 | 20 | 27 |
| C13 | Both | Container CI | 5 | 560 | 0 | 26 | 8 | 28 |
| C29 | Action | Game CI | N/A | 3 | 0 | N/A | 30 | 29 |
| C30 | Both | Backup Utilities | 2 | 22 | 0 | 29 | 27 | 30 |
| C31 | App | Content Att. API | 1 | N/A | 0 | 30 | N/A | 31 |
| C32 | App | GitHub created | 4 | N/A | 0 | 27 | N/A | 32 |

*25% (eight out of 32) of the categories have more than 10 percentage point difference in the percentage of active individuals between researchers and practitioners in the open source community. The researchers are mostly active in 'Code quality' category while practitioners largely focus on 'Continuous integration'.*

## 6.2 The Distribution of Papers and Marketplace Products Across Categories

The distribution of papers and marketplace products vary greatly across categories. For example, 'Continuous Integration' is the #1 and #4 Category for actions and apps, respectively. However, this category contains only five (1.71%) Papers and is ranked #22 out of 27 Categories with at least one paper. On the other hand, 20.89% (61) Papers in our mapping study fall under the 'Testing' category, while only 5.92% (26) Apps and 8.44% (662) Actions on the marketplace fall under this category (see Table 8).

We calculated the percentage of the papers and the marketplace products (apps and actions) in each category and plotted them in Figure 19. In particular, there are 11 Categories with a higher percentage of marketplace products compared to the percentage of papers. This difference is more than 15 Percentage Point (pp) for categories 'Continuous integration' (30.03 pp), 'Deployment' (19.41 pp), and 'Utilities' (16.89 pp). In contrast, 'Code quality' (22.01 pp), 'AI Assisted' (17.48 pp), 'Support' (14.43), 'Testing' (12.59 pp), 'Code search' (11.82 pp) and

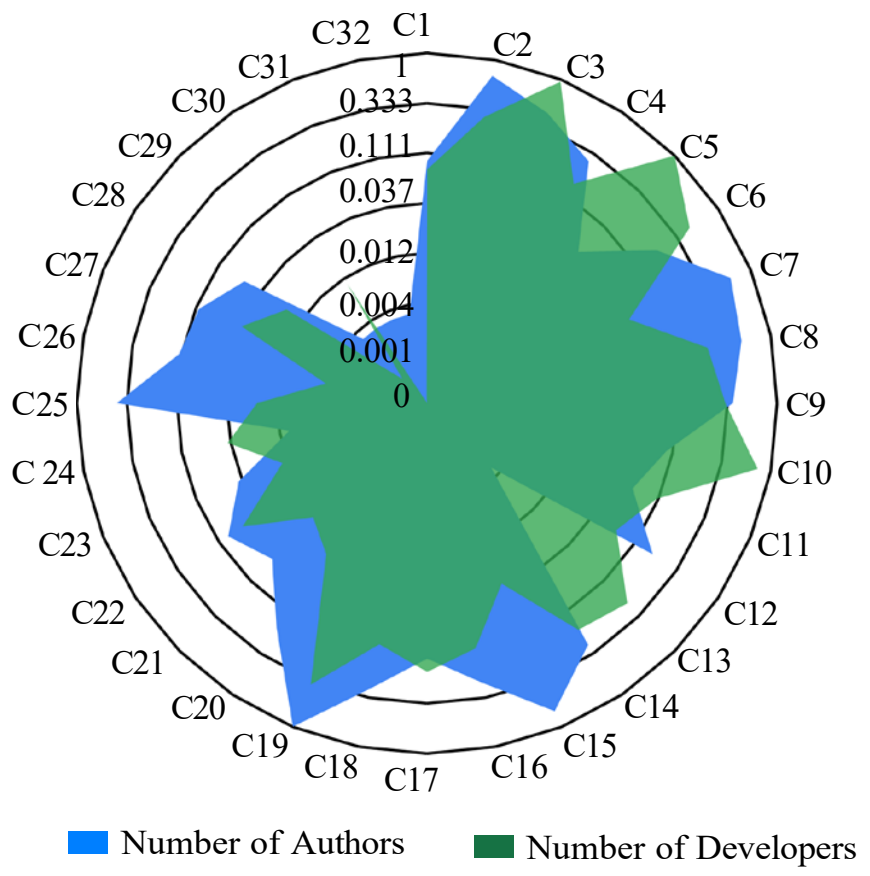

**Fig. 17** Number of Authors vs Number of Developers (after normalizing) per marketplace category

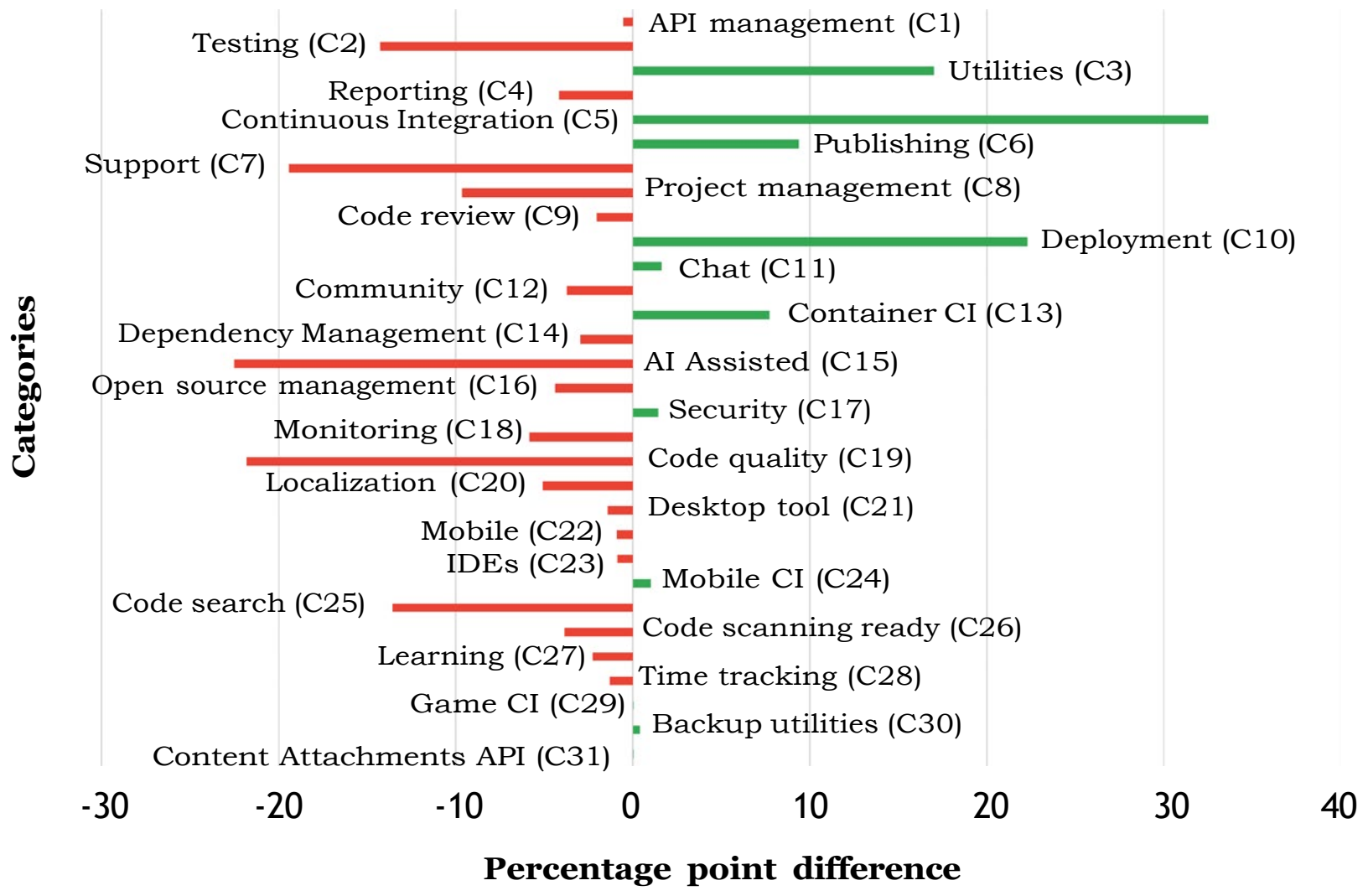

**Fig. 18** Difference (in percentage point) between the number of practitioners and the number of researchers working in each category. Red demonstrates a higher percentage of researchers working in a category, while Green bars indicate a higher percentage of practitioners.

15 other categories have a higher percentage of mapped papers into these categories compared to the percentage of marketplace products.

There were four[22] categories on the GitHub Marketplace for which we did not find any papers within our systematic mapping study, namely 'Container CI', 'Game CI', 'Backup Utilities', and 'Content Attachments API' (see Table 8). Barring these four categories, 'API Management', 'IDEs', 'Mobile', 'Mobile CI', 'Time tracking', and 'Security' had the least percentage point difference (ranging from 0.43 pp to 0.91 pp) between the proportion of the papers and the marketplace products (Figure 19).

*11 categories have a higher percentage of marketplace products compared to 20 categories with a higher percentage of mapped studies. In terms of percentage point difference, literature is lacking the most in 'Continuous integration' while the marketplace falls behind the most in 'Code quality' products. The difference is less that 1 pp for six categories and more than 10 pp for eight other categories.*

## 6.3 Popular Categories in Literature and in the Marketplace

We discussed the interest of researchers and practitioners in developing and contributing to a subject matter in the previous section. Here, we compare the usage and popularity of these topics. Figure 20 shows a bump chart that compares the rankings of categories based on the average number of paper citations, app installs, and action stars. The more popular categories in each column are shown on top while the least popular categories are on the bottom.

[22]excluding 'GitHub Created' which we do not consider for the gap analysis for aforementioned reason

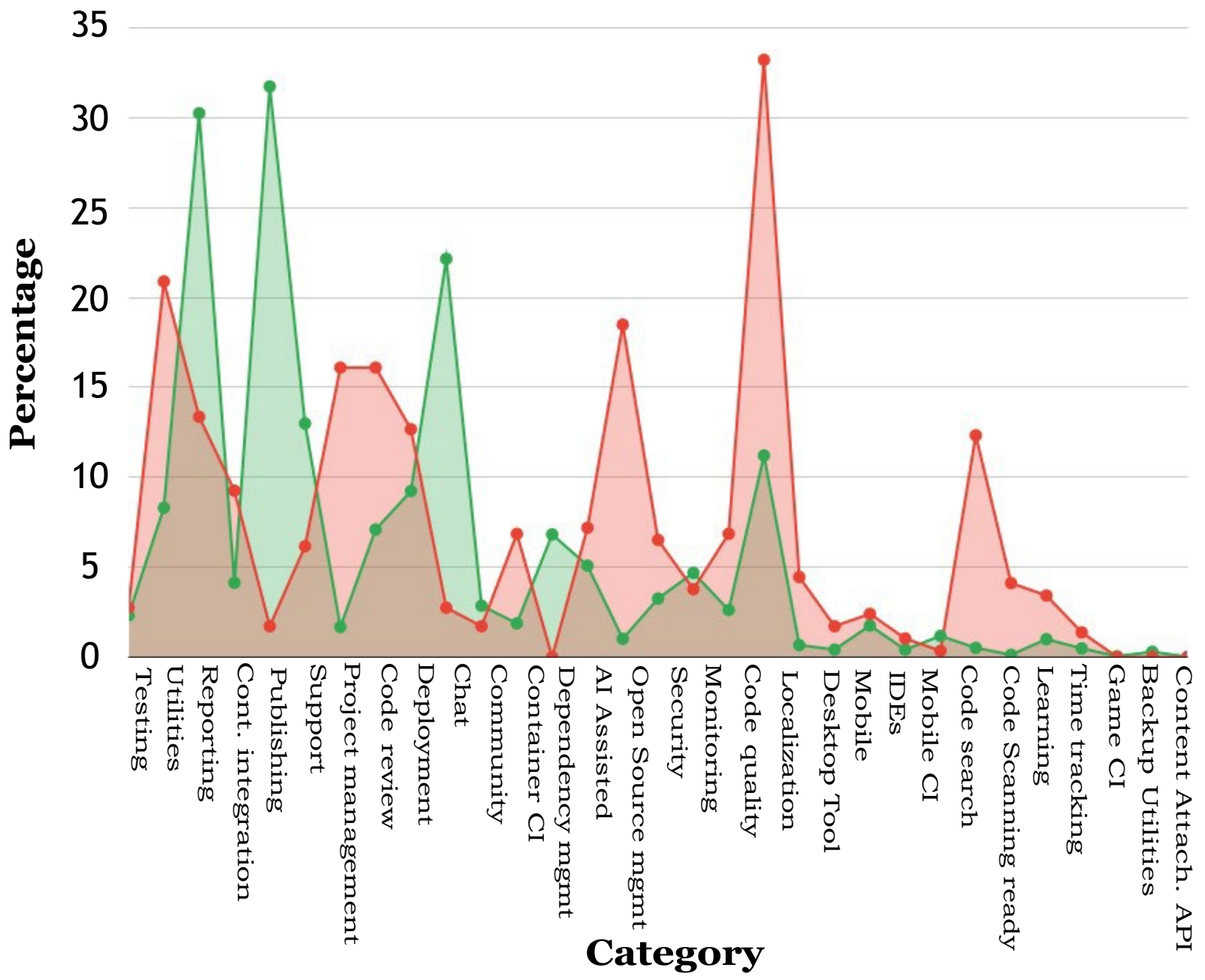

**Fig. 19** The percentage of the marketplace products (green) and papers (red) in each category.

The category 'Mobile CI' displays the most striking gap in popularity between researchers and practitioners. It is ranked $2^{nd}$ and $3^{rd}$ in terms of average installs and stars, respectively, while appearing at the $26^{th}$ position in the average citation ranking. This is because we categorized only one paper in 'Mobile CI' compared to six apps and 91 Actions in this category in the marketplace. 'Time tracking' is another category that is significantly less popular among researchers (#23) compared to practitioners (#1 among action users and #5 among the app users). 'Container CI', 'Desktop tools', 'Security', and 'Monitoring' are other categories that are more favored by practitioners when compared to researchers. On the other hand, categories 'Testing', 'Code search', 'AI Assisted', 'Support', and 'IDEs' are more popular among researchers than both app and action users. The category 'Game CI' ranks #30 (out of 31) in all three metrics, making it one of the least popular categories both in research and in practice. 'Content Attachments API' (#31) and 'Backup Utilities' (#29) are also unpopular among researchers and action users although the categories rank $9^{th}$ and $19^{th}$, respectively, in terms of average number of installs.

*Among categories that are more popular with practitioners, 'Mobile CI' shows the biggest gap compared to popularity with researchers. On the flip side, 'Testing' is more popular among researchers with the largest difference compared to popularity among practitioners. 'Game CI' is equally less popular among both researchers and practitioners.*

### 6.4 Overlap Between Categories in Literature and in the Marketplace

As discussed in section 4.2.2, apps and actions in the marketplace can be listed in more than one category. We looked into how these categories pair up in the marketplace compared to our own categorization of the open-source software engineering literature.

Most frequently, ‘Continuous integration’ and ‘Deployment’ appear together in 7.51% (622 out of 8,285) Products in the marketplace, followed by ‘Continuous integration’ and ‘Utilities’ which are shared by 5.65% (468 out of 8,285) Products. However, in literature we found no paper that falls under both ‘Continuous integration’ and ‘Deployment’. In other words, while there are products in the marketplace that are offering automation both for continuous integration and deployment, research is lacking at the intersection of these two fields. On the other hand, ‘Code quality’ and ‘Code search’ are most frequently shared by studies in literature, appearing together in 8.22% (24 out of 292) Papers (see Figure 16). In contrast, there is no app in the marketplace that falls under both of these categories, and only 0.04% Actions share these two categories.

Figure 21 shows the top three intersections of categories in the marketplace and in the literature. The categories ‘Continuous integration’, ‘Utilities’, and ‘Deployment’ make up the top three pairs of categories shared among the marketplace products. These three categories are also the top three categories based on the number of actions per category. Similarly, the three categories that feature in the most frequently shared categories in the literature- ‘Code quality’, ‘Code review’, and ‘Code search’ are ranked #1, #7, #8, respectively, based on the number of papers per category (see Table 8).

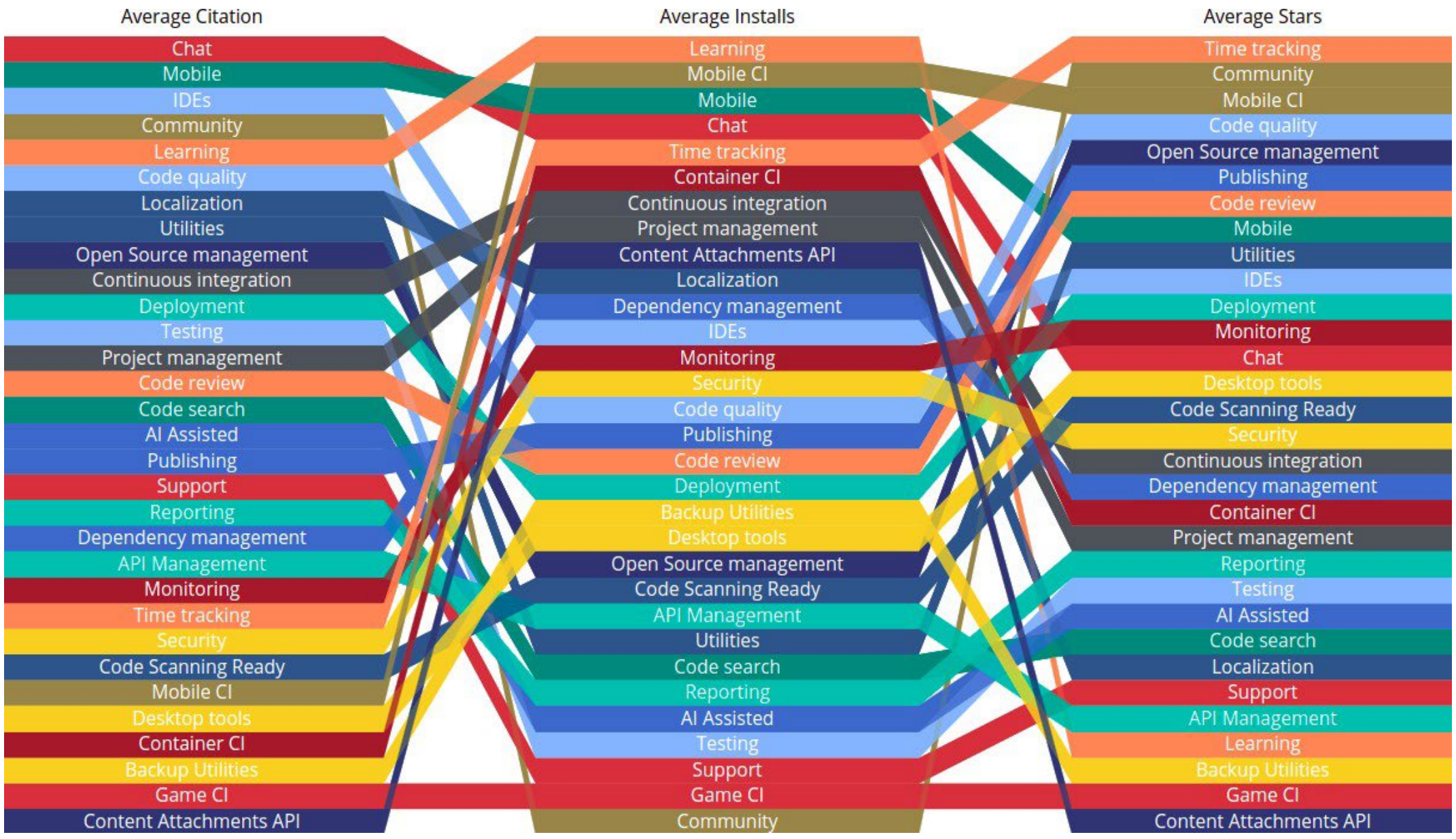

**Fig. 20** Ranking based on the average number of paper citations, app installs, and action stars per category

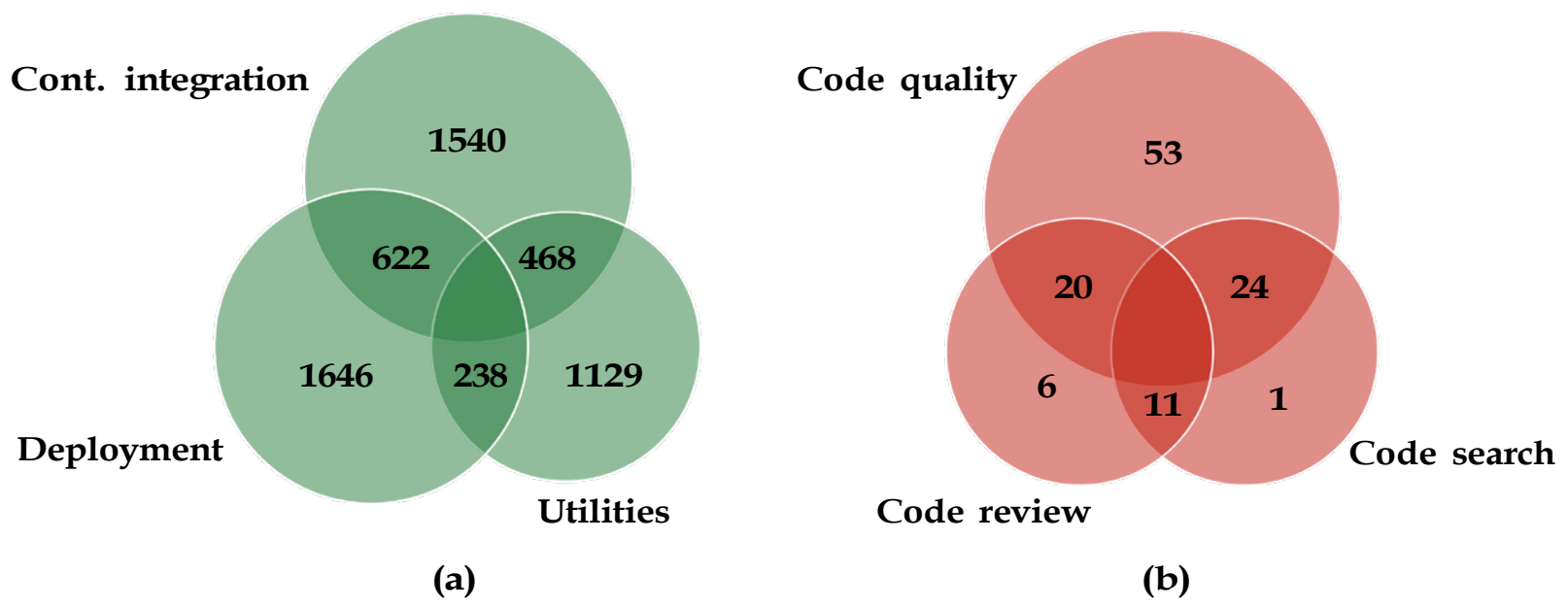

**Fig. 21** The intersection between categories are not equal in the state of the art and state of practice. The left Venn diagram (green) shows the largest overlaps between the marketplace categories (**RQ1**). The right diagram (red) shows the biggest overlaps between the mapped studies (**RQ2**).

*The overlap between categories vary greatly in the marketplace and in literature. 'Continuous integration' and 'Deployment' has the largest intersection in marketplace (appearing together 622 Times) while never overlapping in literature. In contrast, 'Code quality' and 'Code search' overlap frequently in literature (24 Times) but scarcely in the marketplace (3 Times).*

### 6.5 Qualitative Comparison of Automation Tools in Research and Practice

The marketplace products under the 'AI Assisted' category help with code enhancement (3 Mentions in product descriptions) while the studies mapped under this category discuss the task of code generation (5). Code enhancement tasks include improving code readability, fixing bugs in code, adding code comments, etc. On the other hand, code generation helps with producing situation-aware and business aware applications or creating software for embedded systems, among other use cases. For 'API Management', both literature and marketplace products assist with API documentation and API specification. The marketplace offers products under 'Dependency management' that aid developers in finding and resolving dependency conflict (11) as well as updating dependency (6). Similarly, literature in this category propose tools that help with co-changing dependency (4).

The features unit test and test automation are common among both literature and marketplace products that fall under the 'Testing' category. While 'Code Scanning ready' has products that facilitate code scanning (4) compared to papers that propose static analysis (6) and dynamic analysis (6) techniques, they both help with checking security-vulnerability. Products and papers under the 'Publishing' category help developers with their software releases. While the marketplace offers apps and actions that facilitate release creation (266), generating release-notes (178), and release tags (94), the literature discusses more efficient release scheduling (4). Both products and papers offer tools for cloud deployment. 'Code quality' also has products and papers that help with static analysis. As for the 'Code review' category, the marketplace has products for automatically merging pull requests (6), test coverage (5), code linting (64), and a variety of other use-cases. Literature, on the other hand, discuss code analysis (8), program comprehension (14), bug reporting (11), etc under this category.

For 'Community', the marketplace offers products for automatic issue creation and bot comments on issues. On the other hand, the literature discusses using issue reports (4) to identify duplicate issue reports, or to label issues. For example, Di Sorbo et al. [43] analyzed over 6,000 Issue reports in 279 GitHub projects to investigate the nature of 'wontfix issues' and experimented with predicting 'wontfix issues' based on the title and description of issue reports. In the 'Learning' category, marketplace helps with learning skills and sharing knowledge, in comparison to literature which helps with knowledge acquisition (2). In the 'Monitoring' category, there are tools in the marketplace that help with crash reporting or sending notification. The mapped studies focus more on crash localization (3) and crash reproduction (5). For 'Security', marketplace has tools for security analysis, and fixing vulnerability while the literature also discusses code analysis (5) and detecting vulnerability (6).

'Mobile CI' tools on the marketplace are CI tools that are particularly automating the build, test, release, monitoring, and deployment of projects for Android, iOS, Flutter, Firebase, Apache Cordova, Ionic, Octodroid, or Xamarin projects. The marketplace also offers actions for integration with different platforms such as Android emulator, Visual Studio, or Appknox security as well as integration to deploy, label, and upload signed releases to the GooglePlay store or building and publishing APK files on GitHub. We mapped only one paper [42] under this category which discusses automated tests for mobile apps. The authors found that tests for connectivity, GUI, sensors, and multiple configurations were scarce and there was no correlation between automated tests and app popularity.

'Game CI' includes 'tools for building CI pipeline for game development'[23]. Actions in this category are mostly focused on package management and automation to run builds using different criteria or targets. Several of these actions are integrating Unity build engine commands in these processes. In addition, a few of the actions offer notifications on the new releases or set up static data products that host game assets. As for 'Container CI', the GitHub Marketplace offers tools for creating, retrieving, and registering container images across multiple platforms. The actions in this category help with automating tasks such as pushing and running the builds or code files and setting up stream processing platforms such as Apache Kafka and Cordova as well as starting databases. Other than that, actions under 'Continuous integration' offer integration of different tools such as Maven CLI, Armvirt, and Pipenv into GitHub repositories to facilitate developers' activities. We did not find any paper directly related to either 'Game CI' or 'Container CI' within our systematically retrieved list of studies.

The 'Time tracking' category includes apps to extract metadata from developers' activities automatically and use different metrics to monitor or manage projects. The monitoring related apps provide reports on developers' time spent on different projects, using different programming languages, IDEs, workflows, issues, pull requests, commit statuses, and also offer a variety of visualizations or to-do lists. 'Time tracking' tools intersecting with 'Project management' tools are mostly intended to gather billable hours of software developers by tracking the time spent across different desktop and web applications. The actions in this category are being used to block activities such as continuous integration during particular times, posting reminders for different to-do activities, updating README files of projects using stats, providing insight to developers' productivity, automatic decision making on abandoned tasks (such as closing issues which have not been addressed in a specified time period), and sending reports to different tools and mediums. Studies in this category discuss topics such as using git hours to get effort estimations for projects based on git hours per commit, the evolution of projects and artifacts, supporting global software developments by converting time and showing productivity of developers in mornings or nights, automatic tools for adding files changed in pull requests, etc.

[23] https://github.com/marketplace/category/game-ci

## 7 Threats to Validity

This study attempts to obtain a picture of the current state of research and practice in the software engineering community. We have designed measures to elaborate on the state of the art and practice and the mobilization gap. These measures, however, have some threats to validity that we discuss below. We refer to the validity types introduced by Wohlin et al. [50] to structure these limitations.

As for the *conclusion validity*, the threat of drawing wrong conclusions for **RQ1** and **RQ2** are low as the nature of questions are descriptive. We used conventional statistical testing and visualizations to summarize our observations. We used categories as the main unit for our side by side comparison. As we used the marketplace categories and their definitions for our mapping study, we expect these mapping do not pose a huge validity threat. However, the area of categorization is fuzzy and one contribution (either the papers or apps and actions) can be categorized in multiple areas. To provide the readers with further insight, we provided an analysis of the intersection between the categories in Section 6.4.

When it comes to our diagnostic evaluation of the gap between academia and the industry (**RQ3**), there are a number of potential threats. We used the number of papers and authors, as well as the number of apps, actions and their developers as a proxy for the extent of interest from the community. However, these measures might not be an accurate proxy. Moreover, as time passes, the number of citations, in particular, can be affected extensively. Similarly, the popularity (or the perceived value by the user) is inaccurate and only acts as a proxy. To mitigate the risk, we used the average number of citations among all the papers and the average number of installs and stars for apps and actions within each category. These proxies and comparisons might pose a threat on the validity of the conclusions we have drawn.

As per the *construct validity*, the exclusion criteria applied to the systematic mapping may have caused us to ignore some relevant papers in the field. For instance, we collected the papers for our mapping study from the top 20 venues listed in Google Scholar's ranking. This may have excluded some relevant papers from our study. Yet, we believe the search was extensive enough to cover important aspects for our comparison purpose. We also used the categories on GitHub Marketplace to categorize the 365 Papers from our dataset. However, the academic and industry terminology are not consistent. For instance, in the marketplace, 'Localization' is to 'Extend your software's reach. Localize and translate continuously from GitHub.'[24]. While in academia, localization often refers to bug localization. Thus, using just the category name to categorize the research papers and then mapping them to apps and actions on the marketplace could be error-prone. We also rely on the developers to choose appropriate categories for their products on GitHub. However, there could be developers having different perceptions of the categories, making this subjective to the developers' perception. To mitigate this risk, before mapping the studies into the categories in **RQ2**, all the annotators familiarized themselves with the official category definitions in the marketplace. They looked into samples of apps and actions within a category and discussed within the group to have similar understandings of the category definitions.

Further, we used NLP techniques on app and action descriptions to generate word clouds for each category (a sample of word clouds for the 'Localization' and 'Utilities' categories are shown in Figure 22). While this was not good enough for allowing us to automate the categorization process, it gave us a better idea as to what the categories meant in the context of the GitHub Marketplace.

[24] https://github.com/marketplace?category=localization&query=&type=&verification=

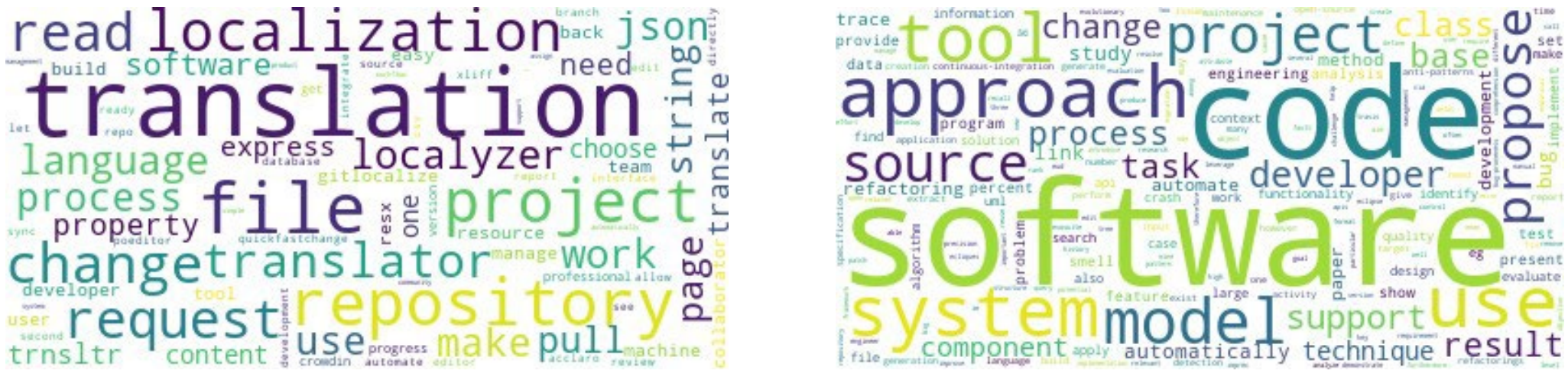

**(a)** **(b)**

**Fig. 22** Word clouds for the categories **(a)** Localization and **(b)** Utilities generated using full descriptions of apps in the categories

When it comes to the *internal validity*, in **RQ2**, we categorized the papers manually. Although the process was independent, after it was done, we discussed the disagreements and used a mediator as to solicit an external opinion and adjusted the categories. Still, there could be bias and misunderstanding that threat the internal validity. It would be ideal to find ways to automate the categorization process, which would make this work free from unintentional research bias. However, as we followed the established protocols of the empirical study and systematic mappings, we expect these biases are minimal.

We also used a combination of an automatic and a manual process to identify individual researchers. We used a combination of names, surnames, and affiliations to identify the individual authors in the field. We followed the process by manually inspecting scholar profiles and LinkedIn accounts. Yet, there is a small chance that some individuals were not identified properly and a few redundancies exist.

As per the *external validity*, for the academic studies (**RQ2**), the publication data and citations over the time was available. However, the marketplace does not provide date for apps and actions over the time (neither in terms of the publication time nor their popularity status over the time). As a result, we could not perform a time based similar analysis or any analysis about evolution of categories over time. This lack of historical data makes it hard to predict upcoming trends in the field. Hence, it is hard to comment on the generalization of the study beyond the scope of this empirical investigation. We expect that in the long run new categories would be introduced or be eliminated from the marketplace.

## 8 Discussion and Implications

To summarize, this paper involves three main contributions.

**First**, we performed the very first study on analyzing the GitHub Marketplace. The study of marketplaces have been of interest in the software engineering community since the emergence of mobile app stores. While there has been intense ongoing research and case studies in the domain of open source software engineering, the marketplace, which is the platform for officially sharing automated tools for the open source ecosystem, has never been studied. By analyzing the marketplace, we believe the research community can invest in mining this repository to gain more insight into different interactions and requirements within the open source software community.

**Second**, we presented the results of a systematic mapping study of the literature in automation within the open source community. While there are multiple studies in the context of open source [25][15][9], ours particularly provides a mapping to the perspective as seen officially by GitHub and the open source developers.

**Third,** we performed a comprehensive comparison between our findings from mining the marketplace and performing the mapping study. This comparison of state of the art and state of the practice demonstrates and quantifies the gap.

GitHub Marketplace is a software repository that *has not* been mined and presented in the software engineering research community. The marketplace is forming a two sided platform (similar to the mobile app stores) where it provides visibility to both app developers and their users [37],[35, 38]. The marketplace provides a combination of qualitative and quantitative data that can be of interest for many in the community. Most interestingly, in contrast to the mobile app stores, not only are the providers software developers but also the users are in software teams and are developers as well.

While this study retrieves information from a new perspective, it should be noted that the community has known and acknowledged the gap between academia and industry. Lo et al. [27] gathered 571 Papers from ICSE and ESEC/FSE venues and asked practitioners to annotate their relevancy. The results showed that for 71% of all ratings, the publications were considered essential or worthwhile while emphasizing on the room to improve relevance and actionability of the research. This study later was replicated by French et al. [14]. On the other hand, Devanbu et al. [11] invested and demonstrated that developers often hold strong a priori opinions about several technical aspects which are not well supported by evidence. Such beliefs formed by subjective understanding and personal experience are error-prone. In a follow up article [12], they echoed the concern of many researchers, "Our life's work, embodied in research papers, counted so little toward forming the opinions of professional practitioners at one of the world's leading software companies?". Further, a more recent study by Shrikanth et al. [41] pointed to the disconnection between beliefs of practitioners and what we achieve with empirical research. The authors advise practitioners and researchers to routinely evaluate their beliefs.

To address the gap, multiple efforts were placed to address the "right question". Buse and Zimmermann [5] empirically evaluated the diverse information needed from stakeholders. Bagel and Zimmermann [4] performed a holistic survey with developers at Microsoft to understand the questions of interest in software teams and listed 145 Questions that could be of interest to the community. Nayebi et al. [35] evaluated the requirements from release engineers for mobile app analysis. These efforts are often employing surveys to understand developers' perceptions. Fernandez et al. [13] performed a large-scale survey with practitioners to identify the pain points of software requirements engineering. These are only a few samples of such surveys to identify the studies relevant to real-world practices. The community's emphasis on performing relevant research for practitioners and lack of access to data from industrial players, open source community, and projects have been the subject of many studies in the field. The research has been looking into the open source software hosted on GitHub. Tools such as GHTotrrent[25] have been designed to facilitate research in the field. We have been hopeful that working on open source has at least gotten us closer to real-world issues. *While many of the studies start with a descriptive analysis of a sample of open source repositories, researchers have not directly evaluated the relevance of research in open source.* Our analysis of **RQ3** does indeed show the gap between the researchers' effort and the practitioners' work in the open source community.

We believe that the marketplace has the potential to help knowledge mobilization if we focus our efforts in that direction. The analysis and mining of the GitHub Marketplace (such as the one done in our study) can *provide researchers with access to a diverse set of perceptions*. A deep look into the developers' ways of automating their tasks on repositories and observing their actions can further reduce the bias of questioning and increase our leverage in narrowing the gap. The *conference tool tracks* can encourage authors who are researching in open source to provide their tools and techniques as an action within the GitHub Marketplace. We think this is a straightforward and compatible way to make a real-world impact and move above silos of research, each performing a case study on a random set of open source repositories. Further, actions being released on the marketplace can also address the issue with replicability and reproducibility of software engineering research.

[25] https://ghtorrent.org

## 9 Conclusion

GitHub Marketplace is a software repository that provides tools for automation based on GitHub events. We mined and analyzed the apps and actions on the GitHub Marketplace and mapped our findings with literature in open source software engineering. The marketplace provides a plethora of information around the tools provided on this platform. These include developers', platform's, and users' usage data on the tools. We also presented a systematic mapping study on literature about automation in open source. We mapped 365 Papers into the categories from the marketplace and summarized our findings. We found that the literature does not cover the automation tools in the GitHub Marketplace very well. Although some of the automation topics in literature are widely used in practice, they have not evolved in a direction to be aligned with the state of practice. We quantified and highlighted the gaps helping researchers and practitioners to identify opportunities and synergies to work toward reducing the gap. There is significant research work on topics such as 'Testing', 'Code quality' and 'Continuous integration' while practitioners are still not making full use of them.

By introducing this software repository, we enabled the community to take further steps in understanding and analyzing different automation tools. The summary of the state of the art helps the researchers to build on top of the existing work while practitioners can navigate the technology and find the synergies. We hope this leads to more vigorous knowledge mobilization within software teams.

## Appendix I